\documentclass[aps,prd,twocolumn,superscriptaddress,showpacs,preprintnumbers,amsmath,amssymb,floatfix]{revtex4-1}

\usepackage[dvipdfmx]{graphicx}
\usepackage{graphicx}
\usepackage{dcolumn}
\usepackage{color}
\usepackage[mathlines]{lineno}
\usepackage{tabularx}
\usepackage{tikz}       
\usepackage{tikz-feynman}
\usepackage{mathtools}
\usepackage{multirow}
\usepackage[titletoc,toc,title]{appendix}
\graphicspath{{ps}}
\usepackage{float}

\definecolor{BlueViolet}{rgb}{0.2, 0.00, 0.7}
\definecolor{Blue}{rgb}{0.15, 0.00, 0.9}
\usepackage[colorlinks=true, linkcolor=Blue, citecolor=Blue, urlcolor=BlueViolet]{hyperref} 
\renewcommand{\arraystretch}{1.1}
\newcolumntype{Y}{>{\centering\arraybackslash}X}

\def\babar{\mbox{\sl B\hspace{-0.4em} {\small\sl A}\hspace{-0.37em} \sl B\hspace{-0.4em} {\small\sl A\hspace{-0.02em}R}}}

\begin{document}
\title{Revisiting fits to $B^{0} \to D^{*-} \ell^{+} \nu_{\ell}$ to measure $|V_{cb}|$ with novel methods and preliminary LQCD data at non-zero recoil}
\author{Daniel Ferlewicz}\affiliation{School of Physics, University of Melbourne, Australia} 
\author{Phillip Urquijo}\affiliation{School of Physics, University of Melbourne, Australia}
\author{Eiasha Waheed}\affiliation{High Energy Accelerator Research Organization (KEK), Japan}

\begin{abstract}
We present a study of fits to exclusive $B^{0} \to D^{*-} \ell^{+} \nu_{\ell}$ measurements for the determination of the Cabbibo-Kobayashi-Maskawa matrix element magnitude $|V_{cb}|$, based on the most recent Belle untagged measurement. 
Results are obtained with the Caprini-Lellouch-Neubert (CLN) and Boyd-Grinstein-Lebed (BGL) form factor parametrizations, with and without the inclusion of preliminary Lattice QCD measurements of form factors at non-zero hadronic recoil from the JLQCD collaboration. The CLN and BGL fits are also studied in different scenarios with reduced theoretical assumptions, and at higher order expansions, respectively. To avoid bias from high systematic error correlations we employ a novel technique in the field of $B$-physics phenomenology with a toy MC using a Cholesky decomposition of the covariance matrix. 
Using additional input from Lattice QCD calculations of form factors at non-zero recoil, in collaboration with JLQCD, allows for well-defined fit results with reduced model dependence in CLN and BGL.
The results obtained are consistent between different configurations, ultimately providing a method for a more model-independent exclusive measurement of $|V_{cb}|$. Using preliminary inputs, $\mathcal{F}(1)\eta_{\rm EW}|V_{cb}|$ is found to be approximately $(35.02 \pm 0.29 \pm 0.88) \times 10^{-3}$ in BGL(2,2,2) and $(34.96 \pm 0.32 \pm 0.96) \times 10^{-3}$ in CLNnoHQS.
\end{abstract}

\maketitle

\section{Introduction}
\label{sec:Intro}
The Cabbibo-Kobayashi-Maskawa matrix element $|V_{cb}|$ is a fundamental parameter of the Standard Model, describing the weak decay of $b$-quarks and must be measured. A long-standing discrepancy between inclusive and exclusive decay mode determinations limits our present understanding of this parameter.
The world-average value from combined inclusive results is based on measurements of semileptonic $B$ meson decays of the type $B \to X_c\ell \nu$, where $X_c$ denotes all possible hadronic states in the $b \to c \ell \nu$ transition. The inclusive value is reported to be~\cite{cite-2020PDG}
\begin{eqnarray}
|V_{cb}| &=& (42.2 \pm 0.8) \times 10^{-3}.
\end{eqnarray}
The exclusive measurements are determined from both $B^{0} \to D^{*-}\ell^+ \nu_\ell$ and $B^{0} \to D^{-}\ell^+ \nu_\ell$ decays. The results from $B^{0} \to D^{*-}\ell^+ \nu_\ell$ decays are
\begin{eqnarray}
|V_{cb}| &=& (38.4 \pm 0.7 \pm 0.5 \pm 1.0) \times 10^{-3},
\label{eq:pdg_excl}
\end{eqnarray}
using ${\cal F}(1) = 0.904 \pm 0.012$, where the first uncertainty is experimental, the second is from Lattice QCD, and the third is an additional uncertainty to compensate for the truncation of power series in a fit. The results from $B \to D \ell^+ \nu_\ell$ decays are
\begin{eqnarray}
|V_{cb}| &=& (40.1 \pm 1.0) \times 10^{-3},
\end{eqnarray}
using ${\cal G}(1) = 1.054 \pm 0.004 \pm0.008$. Both exclusive results are significantly lower than that from the inclusive approach~\cite{cite-2020PDG}. These modes must be studied in further detail to understand where the discrepancies could be originating.

Measurements of $|V_{cb}|$ from $B^{0} \to D^{*-}\ell^+ \nu_\ell$ decays have typically used two different parametrizations for models in order to calculate these extrapolations; one by Caprini, Lellouch and Neubert (CLN) \cite{cite-dispersivebounds} and another by Boyd, Grinstein, and Lebed (BGL) \cite{cite-Boyd}. Both the CLN and  BGL parametrizations are built from the same foundations of operator product expansions, where analytic properties of form factors can also introduce some constraints. The CLN parametrization makes use of Heavy Quark Effective Theory (HQET) relations and its constraints on the $B^{0} \to D^{*-}\ell^+ \nu_\ell$ form factors to reduce the number of independent free parameters with respect to BGL. 

The 2019 untagged Belle measurement of $|V_{cb}|$ \cite{cite-Eiasha} measured binned yields of $B^{0} \to D^{*-}\ell^{+} \nu_{\ell}$ and fits to the CLN and BGL parametrizations, measuring $\mathcal{F}(1)\eta_{\rm EW}|V_{cb}| = (35.06 \pm 0.15 \pm 0.56) \times 10^{-3}$ and $(34.93 \pm 0.23 \pm 0.59) \times 10^{-3}$, respectively, where the latter is determined at a fixed order of power series truncation, defined later. These results showed consistency between the two parametrizations at a given order in the BGL expansion, in contrast to the results based on a preliminary measurement with tagged Belle data \cite{Abdesselam:2017kjf}, as covered by Refs. \cite{Grinstein:2017nlq,Bigi:2017njr} and others. Both fit studies acknowledge that the observed discrepancy may have been a feature of the tagged dataset. The Belle results are consistent with a four-dimensional analysis from \babar{} \cite{Dey:2019bgc}, with $\mathcal{F}(1)\eta_{\rm EW}|V_{cb}| = (35.02 \pm 0.77) \times 10^{-3}$ in CLN and $(34.98 \pm 0.82) \times 10^{-3}$ in BGL, where the BGL parametrization also required truncation. However, various studies \cite{GAMBINO2019386,Bernlochner:2019ldg} have shown that the choice of configuration in CLN and BGL and the implementation of systematic uncertainties has an effect on the measured value of $|V_{cb}|$ and the relevant form factors in exclusive measurements. In this study, we use both the CLN parametrization and the BGL parametrization in fits to further explore the Belle results, checking for consistency in measurements between parametrizations and the effects of systematic correlations. We apply a technique in which the Cholesky decomposition of a covariance matrix is used to generate a toy Monte Carlo sample in a novel approach for fits to $B$-meson decays. We then include additional data from preliminary LQCD calculations to allow for fits to higher order in the BGL parametrization and to explore fits with different theoretical assumptions for CLN that are less model-dependent.

This paper is organized as follows. Section \ref{sec:Overview} is a description on the techniques used to extract results from the data provided by the 2019 Belle measurement, followed by a brief summary of the conventions used in this analysis. In Section \ref{sec:Cholesky}, we discuss a method in which systematic uncertainties can be taken into account when correlations between observed measurements are very high (predominantly from scale uncertainties). 
Sections \ref{sec:CLNs} and \ref{sec:BGLs} explore fits to the CLN and BGL parametrizations with more degrees of freedom, followed by the incorporation of additional LQCD data in Section \ref{sec:LQCD}. The results are then summarized and discussed in Section \ref{sec:Conc}.

\section{Analysis overview}
\label{sec:Overview}
The exclusive value of $|V_{cb}|$ is typically extracted from fits to yields of $B^{0} \to D^{*-}\ell^+ \nu_\ell$ decays as a function of the kinematic observables: the hadronic recoil, $w$, and three angular variables $\cos{\theta_\ell}$, $\cos{\theta_v}$ and $\chi$. Hadronic recoil is defined as
\begin{equation}
    w = \frac{m_{B^{0}}^2 +m_{D^{*\pm}}^2-q^2}{2m_{B^{0}}m_{D^{*\pm}}},
\end{equation}
where $q^2$ is the invariant mass squared of the lepton-neutrino system and $m_{B^{0}}$ and $m_{D^{*\pm}}$ are the $B^{0}$ and $D^{*\pm}$ meson masses, respectively. The observable $\theta_{\ell}$ is the angle between the direction of the lepton and the direction opposite to the $B$ meson in the $W$ boson rest frame and $\theta_{v}$ is the angle between the direction of the $D^{0}$ meson and the direction opposite to the $B$ meson in the $D^{*}$ meson rest frame. The observable $\chi$ is the angle between the planes formed by the decays of the $W$ and $D^{*}$ mesons in the rest frame of the $B$ meson.
The analysis described here follows the approach taken in the 2019 Belle analysis~\cite{cite-Eiasha}, where a theoretical differential decay rate is calculated in 10 bins for each of the observables and then forward-folded with the detector response. A partial integration of the full four-dimensional differential decay rate,
\begin{eqnarray}\label{eq:diff}
&&\frac{d\Gamma (B^{0} \to D^{*-}\ell^+ \nu_\ell)}{dwd\cos\theta_\ell d\cos\theta_Vd\chi} = \nonumber\\
&& \frac{\eta_{\rm EW}^2 3 m_{B^{0}} m^2_{D^{*\pm}}}{4(4\pi)^4} G_F^2 |V_{cb}|^2\sqrt{w^2-1}(1 - 2wr + r^2)   \nonumber \\
&&\left\{ ( 1 - \cos\theta_\ell)^2\sin^2\theta_V H_+^2 + (1+\cos \theta_\ell)^2\sin^2 \theta_V H_-^2  \right. \nonumber \\
&& +4\sin^2 \theta_\ell \cos^2 \theta_V H_0^2 - 2\sin^2 \theta_\ell \sin^2 \theta_V\cos2\chi H_+H_-  \nonumber \\
&& -4\sin\theta_\ell(1-\cos\theta_\ell)\sin\theta_V\cos\theta_V\cos\chi H_+H_0 \nonumber \\
&& \left. +4\sin\theta_\ell(1+\cos\theta_\ell)\sin\theta_V\cos\theta_V\cos\chi H_-H_0 \right\},
\end{eqnarray}
is performed to obtain an expression for the decay width in terms of the three helicity amplitudes associated with this decay, $(H_{\pm},H_0)$, as a function of a set of model parameters, $\mathbf{x}$. The helicity amplitudes are defined in the following section, $G_F$ is the Fermi constant and $\eta_{\rm EW}$ is an electroweak correction for the semileptonic decay \cite{Sirlin:1981ie}. These calculations are then integrated over a fixed bin width for each observable to obtain the vector for the unfolded integrated yield, $N^{\rm int.}(\mathbf{x})$, before a detector response matrix, $R$, is applied in combination with the efficiency for reconstructing an event $\epsilon$, to obtain an expected yield, $N^{\rm exp.}(\mathbf{x})$, in each bin, $i$:
\begin{equation}
    N_i^{\rm exp}(\mathbf{x}) = \sum_{j=1}^{40} R_{ij}\epsilon_j N_{j}^{\rm int.}(\mathbf{x}).
\end{equation}
The expected values are compared to the data in each bin with a $\chi^2$ minimization algorithm to determine the most likely set of parameters $\mathbf{x}$ that will model the data.

The fit is used to extrapolate measurements of the form factors for this decay to the zero-recoil point, $w=1$, where the model parameters are described and constrained by unquenched Lattice Quantum Chromodynamic (LQCD) calculations. 

\section{Form Factor parametrization}\label{Section-FF}
In the standard CLN parametrization, the three helicity amplitudes are defined as follows:
\begin{equation}
    H_i(w) = m_{B^{0}} \frac{r' (1-r^2)(w+1)}{2 \sqrt{1 - 2wr + r^2}} h_{A_1}(w) |\tilde{H}_i(w)|,
\end{equation}
and
\begin{align}
    \tilde{H}_\pm &= \frac{\sqrt{1 - 2wr + r^2} (1 \mp \sqrt{\frac{w-1}{w+1}}R_1(w))}{1-r}, \nonumber\\
    \tilde{H}_0 &= 1 + \frac{(w-1)(1-R_2(w))}{1-r},
\end{align}
where $r = m_{D^{*\pm}}/m_{B^{0}}$ and $r' = 2 \sqrt{m_{B^{0}} m_{D^{*\pm}}}/(m_{B^{0}} + m_{D^{*\pm}})$.

The form factor $h_{A_1}$ and the form factor ratios $R_1$ and $R_2$ are defined in terms of the free parameters $\rho^2$, $R_1(1)$ and $R_2(1)$ as 
\begin{eqnarray}\label{eq:cln}
h_{A_1}(w) &=& h_{A_1}(1)\left[ 1 - 8 \rho^2z(w) + (53\rho^2 - 15)z(w)^2  \right. \nonumber\\
&&\left. - (231\rho^2-91)z(w)^3\right], \nonumber\\
R_1(w) &=& R_1(1) - 0.12 (w-1) + 0.05 (w-1)^2, \nonumber\\
R_2(w) &=& R_2(1) + 0.11 (w-1) - 0.06 (w-1)^2, 
\end{eqnarray}
where $z(w) = \frac{(\sqrt{w+1}-\sqrt{2})}{(\sqrt{w+1}+\sqrt{2})}$. 
These are combined to obtain an overall expression for he form factors:
\begin{eqnarray}
{\mathcal F}^2(w) =&& h_{A_1}^2(w) \Big( 1 + 4 \frac{w}{w+1} \frac{1- 2wr+r^2}{(1-r^2)}\Big)^{-1} \nonumber\\
&&\Big[ 2 \frac{1- 2wr+r^2}{(1-r)^2} \left( 1 + R_1^2(w)\frac{w-1}{w+1}\right) \nonumber\\
&& + (1 + (1-R_2(w)) \frac{w-1}{1-r})^2\Big].
\end{eqnarray}
Perfect heavy quark symmetry, in the limit of infinite quark mass, implies equality between all form factors and ratios, $\mathcal{F} (w) = R_{1}(w) = R_{2}(w) = 1$. The finite masses of quarks can then be accounted for in corrections at zero hadronic recoil, resulting in $h_{A_{1}}(1) = \mathcal{F} (1) = 0.906$ \cite{Bailey:2014tva}. 
Therefore, there are four independent parameters in this model used to calculate the expected yield of $B^{0} \to D^{*-}\ell^{+} \nu_{\ell}$ events: $|V_{cb}|$, $\rho^2$, $R_{1}(1)$ and $R_{2}(1)$. The values of these parameters are not calculated, but are instead extracted from fits to experimental data. We note that isolating the value of $|V_{cb}|$ from fits depends on knowing the scale factor $\mathcal{F} (1)$ to a high degree of certainty. As this number is subject to change, we present all results for $|V_{cb}|$ in the form $\mathcal{F}(1)\eta_{\rm EW}|V_{cb}|$, except where cancellation occurs in ratios.

In the BGL parametrization, the helicity amplitudes are defined as
\begin{align}
    H_0(w) &= \mathcal {F}_1(w)/\sqrt{q^2}, \nonumber\\
    H_\pm(w) &= f(w) \mp m_{B^{0}} m_{D^{*\pm}} \sqrt{w^2 -1} g(w),
\end{align}
where three power series
\begin{eqnarray}\label{eq:bglff}
f(z) &=& \frac{1}{P_{1+}(z)\phi_f(z)}\sum_{n=0}^{\infty}a_n^fz^n~,\nonumber\\
\mathcal {F}_1(z) &=& \frac{1}{P_{1+}(z)\phi_{\mathcal F_1}(z)}\sum_{n=0}^{\infty}a_n^{{\mathcal F_1}}z^n~,\nonumber\\
g(z) &=& \frac{1}{P_{1-}(z)\phi_g(z)}\sum_{n=0}^{\infty}a_n^gz^n~,
\end{eqnarray}
are related to the CLN form factors via 
\begin{eqnarray}\label{eq:cln-eqs}
h_{A_{1}}(w)&=& \frac{f(w)}{\sqrt{m_{B^{0}}m_{D^{*\pm}}} (1+w)},\nonumber\\
h_V(w)&=& g(w)\sqrt{m_{B^{0}}m_{D^{*\pm}}},\nonumber\\
R_{1}(w)& =& (w+1)m_{B^{0}}m_{D^{*\pm}}\frac{g(w)}{f(w)},\nonumber\\
R_{2}(w)& =& \frac{w-r}{w-1} - \frac{\mathcal{F}_1(w)}{m_{B^{0}} (w-1)f(w)}.
\end{eqnarray}
In these equations, the Blaschke factors, $P_{1\pm}$, are given by
\begin{eqnarray}
P_{1\pm}(z) &=& \prod_{P=1}^{n} \frac{z-z_{\pm P}}{1-zz_{\pm P}}~,
\end{eqnarray}
where $z_{\pm P}$ is defined as
\begin{eqnarray}
z_{\pm P} &=& \frac{\sqrt{t_+ - m_{\pm P}^2}- \sqrt{t_+-t_-} }{\sqrt{t_+ - m_{\pm P}^2} + \sqrt{t_+-t_-} }~.
\end{eqnarray}
Here $t_{\pm}=(m_{B}\pm m_{D^{*}})^2$ and $m_{\pm P}$ denotes the $P^{\rm th}$ mass of the $n$ $B_c^*$ $1^{\pm}$  resonances available (see Table~\ref{tab:constants}). The functions $\phi_i(z)$ are outer functions related to these Blaschke factors \cite{cite-Eiasha}. 
We have adopted the notation from Ref. \cite{GAMBINO2019386}, where \textbf{$(n_f, n_g, n_{\mathcal{F}_1})$} refers to the highest power in each of these series that has not been fixed to zero. The 2019 Belle analysis used the BGL(1,0,2) configuration, with five free parameters. This is due to instability ($i.e.\ $ either a lack of convergence in fits using MINUIT~\cite{James:1994vla}, parameters being returned as values within $1\sigma$ of their bounds, or from poorly defined or non-unique $\chi^2$ minima) when more parameters were included. Through the inclusion of Lattice QCD inputs, we will explore higher order expansions in this study and test for stability in $|V_{cb}|$ as the number of free parameters is increased. 

Unitarity constraints on the series coefficients require~\cite{cite-Boyd}
\begin{eqnarray}\label{eq:eqs-constraints}
\sum_{n=0}^{\infty} (a_n^g)^2 <1~,\nonumber\\
\sum_{n=0}^{\infty} \left[ (a_n^f)^2 + (a_n^{\mathcal F_1})^2 \right] <1~,
\end{eqnarray}
which have been enforced through a hard cut-off in the $\chi^2$ minimization.
We note that by redefining these coefficients as $\tilde{a} = \eta_{\rm EW}|V_{cb}|a$, we may extract a value for $|V_{cb}|$ using
\begin{eqnarray}
\mathcal{F}(1)\eta_{\rm EW}|V_{cb}| &=& \frac{1}{2m_{B^{0}} m_{D^{*\pm}}}  \frac{|\tilde{a}_0^f|}{P_f(0) \phi_f(0)}.
\label{eq:a0f_to_Vcb}
\end{eqnarray}
A list of the inputs used in this analysis is given in Table~\ref{tab:constants}, with values chosen to remain consistent with Ref.~\cite{cite-Eiasha}, except for $\mathcal{B}(D^{0} \to K^{-} \pi^{+})$, which uses the average result from the 2020 PDG \cite{cite-2020PDG}. There have been recent changes in this value, which has an impact on the normalization of measurements in this study and therefore directly affects the obtained values for $|V_{cb}|$.

\begin{table}[htb]
\centering
\caption{The full set of inputs used in this study. Both the CLN and the BGL fits use the common input, in addition to the data yields and detector responses published in Ref.~\cite{cite-Eiasha}. The uncertainties listed are incorporated into the global systematic covariance matrix.}
 \renewcommand{\arraystretch}{1.3}
\begin{tabularx}{0.99\linewidth}{Y Y} 
\hline \hline
\multicolumn{2}{c}{Common Input} \\
\hline
$\eta_{\rm EW}$ & $1.0066$ \\
$h_{A_1}(1) = \mathcal{F}(1)$ & $0.906$\\
$m_{B^{0}}$ & $5.27963 \pm 0.00015$ GeV/$c^2$\\
$m_{D^{*\pm}}$ & $2.01026 \pm 0.00005$ GeV/$c^2$\\
$\tau_{B^{0}}$ & $(1.520 \pm 0.004) \times 10^{-12}$ s \\
$\mathcal{B}(D^{*+} \to D^{0} \pi^{+})$ & $0.677 \pm 0.005$\\
$\mathcal{B}(D^{0} \to K^{-} \pi^{+})$ & $0.0391 \pm 0.0003$\\
$G_{F}$ & $1.16637 \times 10^{-5}$ GeV$^{-2} (\hbar c)^3$\\
$2n_{B\bar{B}}f_{00}$ (Belle) & $(750 \pm 11) \times 10^6$\\
\hline
\multicolumn{2}{c}{BGL Input} \\
\hline
$B_c^*$ $1^+$ masses & $6.730$ GeV/$c^{2}$\\
 & $6.736$ GeV/$c^{2}$\\
 & $7.135$ GeV/$c^{2}$\\
 & $7.142$ GeV/$c^{2}$\\
$B_c^*$ $1^-$ masses & $6.337$ GeV/$c^{2}$\\
 & $6.899$ GeV/$c^{2}$\\
 & $7.012$ GeV/$c^{2}$\\
 & $7.280$ GeV/$c^{2}$\\
$n_I$ & 2.6 \\
$\chi^T(+u)$ & $5.28 \times 10^{-4}$ (GeV/$c^{2})^{-2}$\\
$\chi^T(-u)$ & $3.07 \times 10^{-4}$ (GeV/$c^{2})^{-2}$\\
\hline 
\hline
\end{tabularx}
\label{tab:constants}
\end{table}

\section{The Cholesky decomposition method}
\label{sec:Cholesky}
A common method of measuring free parameters is via a $\chi^2$ fit, maximizing the likelihood of observing an obtained binned dataset.
This is achieved by an algorithm that iterates through different values of a set of parameters, $\mathbf{x}$, to minimize a $\chi^2$ variable, defined typically as
\begin{eqnarray}\label{eq:chi2}
\chi^2 &=& \sum_{i,j}^{} \left(N_i^{\rm obs} - N_i^{\rm exp} \right)\mathcal{C}^{-1}_{ij} \left( N_j^{\rm obs} - N_j^{\rm exp} \right),
\end{eqnarray}
where $N_i^{\rm obs}$ is the number of events observed in bin $i$ of the data sample, $N_i^{\rm exp}$ is the number of events expected in bin $i$, determined from theory using $\mathbf{x}$, and $\mathcal{C}^{-1}$ is the inverse of the covariance matrix. 
The correlations between observed values are related to the covariance matrix by
\begin{eqnarray}
\rho{(i,j)} = \frac{\mathcal{C}_{ij}}{\sigma_i \sigma_j},
\end{eqnarray}
where $\sigma_i$ is the standard deviation of the $i^{\rm th}$ measurement ($i.e.\ $ $\sigma_i^2 = \text{var($i$)} = \mathcal{C}_{ii}$).

In evaluating $\chi^2$, the diagonal elements ($i=j$) must add a positive value, while those off the diagonal may add a negative value. Data samples with large off-diagonal correlations can therefore have a lower $\chi^2$ than an identical data sample with correlations closer to zero. This can place a minimum $\chi^2$ at parameter values away from those which provide the nominal fit ($i.e.\ $the expected events from the fit will appear biased away from the observed data). Bias can be introduced by incorrectly applying multiplicative systematic uncertainties to data yields rather than the theoretical expectation~\cite{Jung:2018lfu}, known as the D'Agostini effect~\cite{DAGOSTINI}. As in the original Belle analysis, this effect is correctly treated here.

To determine the uncertainty of a parameter in cases where correlations are high we use a toy Monte Carlo (MC) method incorporating the Cholesky decomposition~\cite{cite-chol} to avoid directly including systematic uncertainties from normalization factors into the $\chi^2$. A positive-definite covariance matrix, $\mathcal{C}$, undergoes a Cholesky decomposition such that
\begin{eqnarray}
\mathcal{C} = L L^T,
\end{eqnarray}
where $L$ is a lower triangular matrix. A vector $u$ is created where each element is a random number taken from a Gaussian distribution with a mean of $0$ and variance $1$. The vector for the observed data is then fluctuated by
\begin{eqnarray}\label{chol_eqn}
N'_{obs} = N_{obs} + Lu,
\end{eqnarray}
where $Lu$ maintains the desired covariance properties \cite{Waldron_2012}.
The data is fluctuated many times by resampling the vector $u$, and the fit procedure is repeated to find a distribution of parameter values determined by the $\chi^2$ minimization algorithm. This process represents the randomness in measurements of the data that are affected by systematic or statistical uncertainties and correlations incorporated in the covariance matrix. A Gaussian function can then be fit to these parameter distributions and the mean and width can be taken to be the nominal value and uncertainty of each parameter, respectively. Assuming that all fits converged without reaching boundary conditions, the mean obtained from the fit without the toy MC method will be equal to the mean obtained from the Gaussian fit. Likewise, the standard deviation calculated from the distribution of results will be equal to the width obtained from the Gaussian fit and it is through this latter method that central values and uncertainties are calculated in this analysis. 

\subsection{Using the Cholesky decomposition method}
In the Belle paper \cite{cite-Eiasha}, the systematic uncertainties were determined through a toy MC approach and were published as a systematic uncertainty correlation matrix. Although the statistical correlations between the bins of the different observables were small, the systematic correlations were close to 1 almost everywhere as they were dominated by scale errors. A fit using the data provided obtains the same central values when only the statistical covariance matrix is used in the $\chi^2$ function. However, if systematic uncertainty is introduced into the covariance matrix in a ``naive" way, where 
\begin{equation}
\mathcal{C}_{\rm total} = \mathcal{C}_{\rm stat.} + \mathcal{C}_{\rm sys.},
\end{equation}
the minimizing algorithm returns a lower $\chi^2$ value but with a clear bias in the expected yield and fit parameters. These results are discussed in Appendix \ref{appendix:A}.

To perform fits using the data provided in the Belle paper, separate toy MC samples were generated using the Cholesky decomposition method for each of the statistical and systematic covariance matrices. Each sample contained $10^4$ iterations of $N'_{obs}$, which was then used to determine a set of fit parameters by minimizing the $\chi^2$ from Eq.~\ref{eq:chi2}, where $\mathcal{C}$ was the statistical covariance matrix in both samples. We note that due to limits in numerical precision of the data available, the systematic covariance matrix requires a slight correction to maintain positive-definiteness; this is done by multiplying the elements along the diagonal by a factor of $(1+10^{-6})$.

\begin{figure*}[htb]
    \centering
    \begin{minipage}{0.4\linewidth}
    \includegraphics[width=\linewidth]{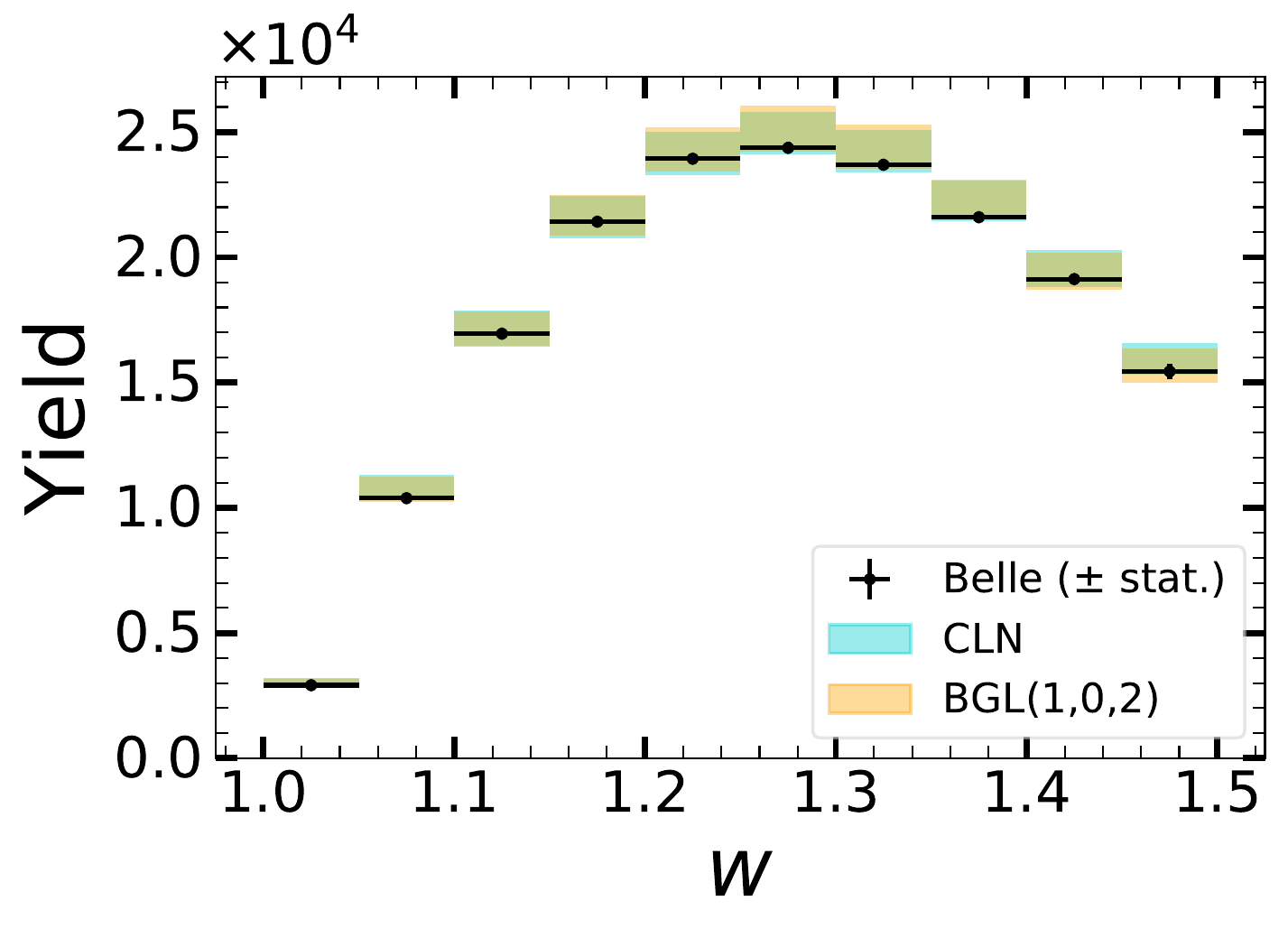}
    \includegraphics[width=\linewidth]{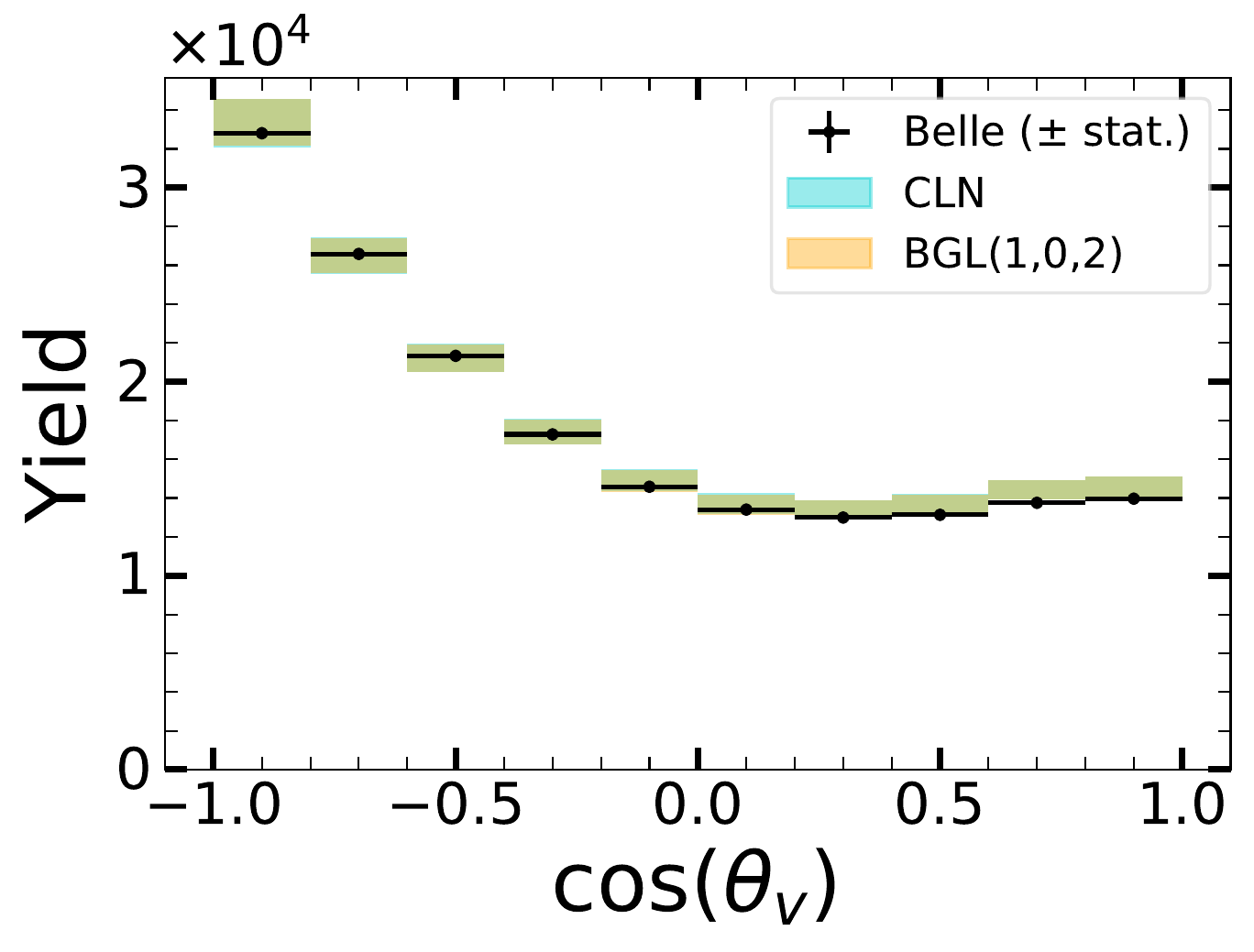}
    \end{minipage}
    \begin{minipage}{0.4\linewidth}
    \includegraphics[width=\linewidth]{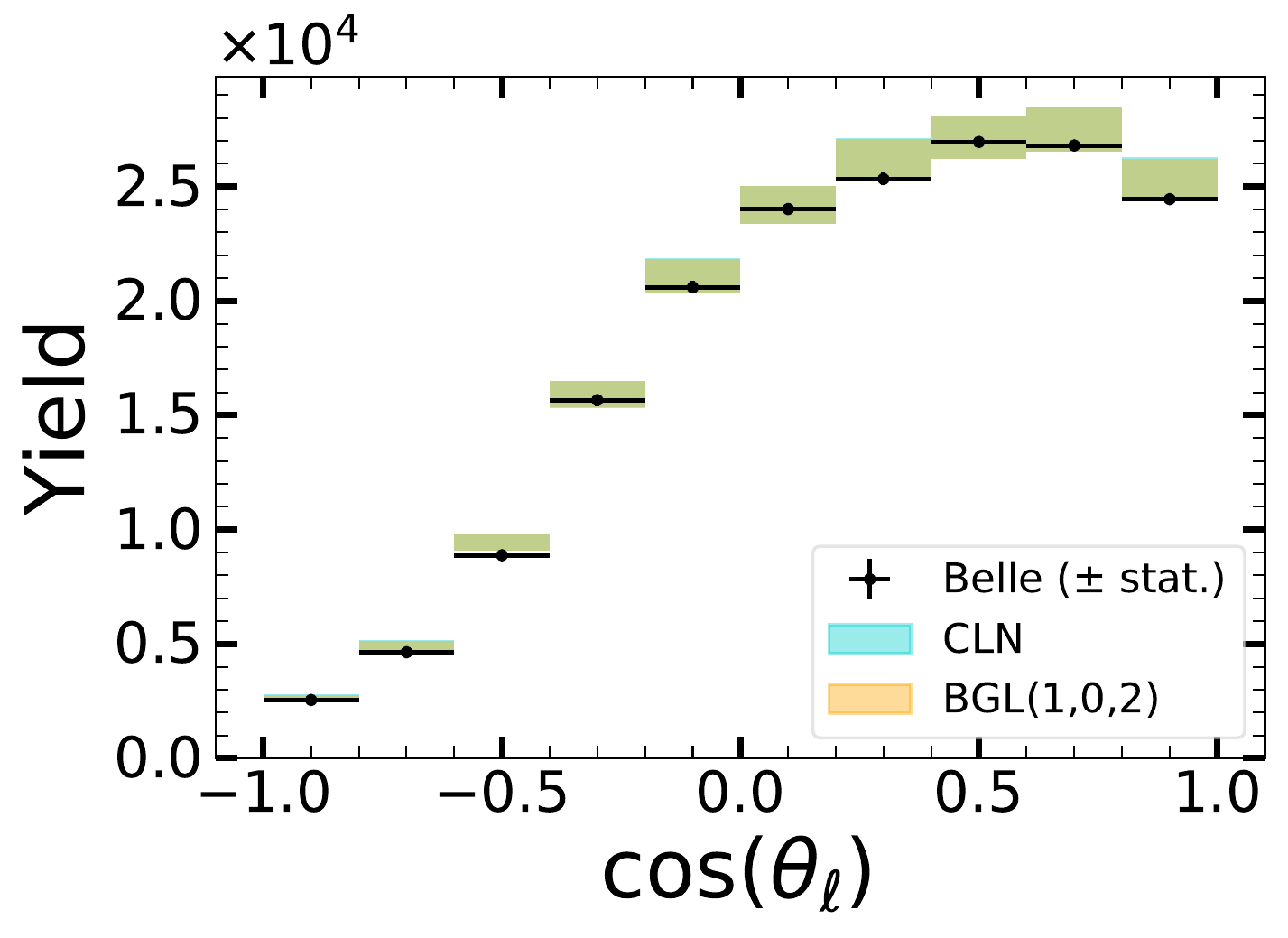}
    \includegraphics[width=\linewidth]{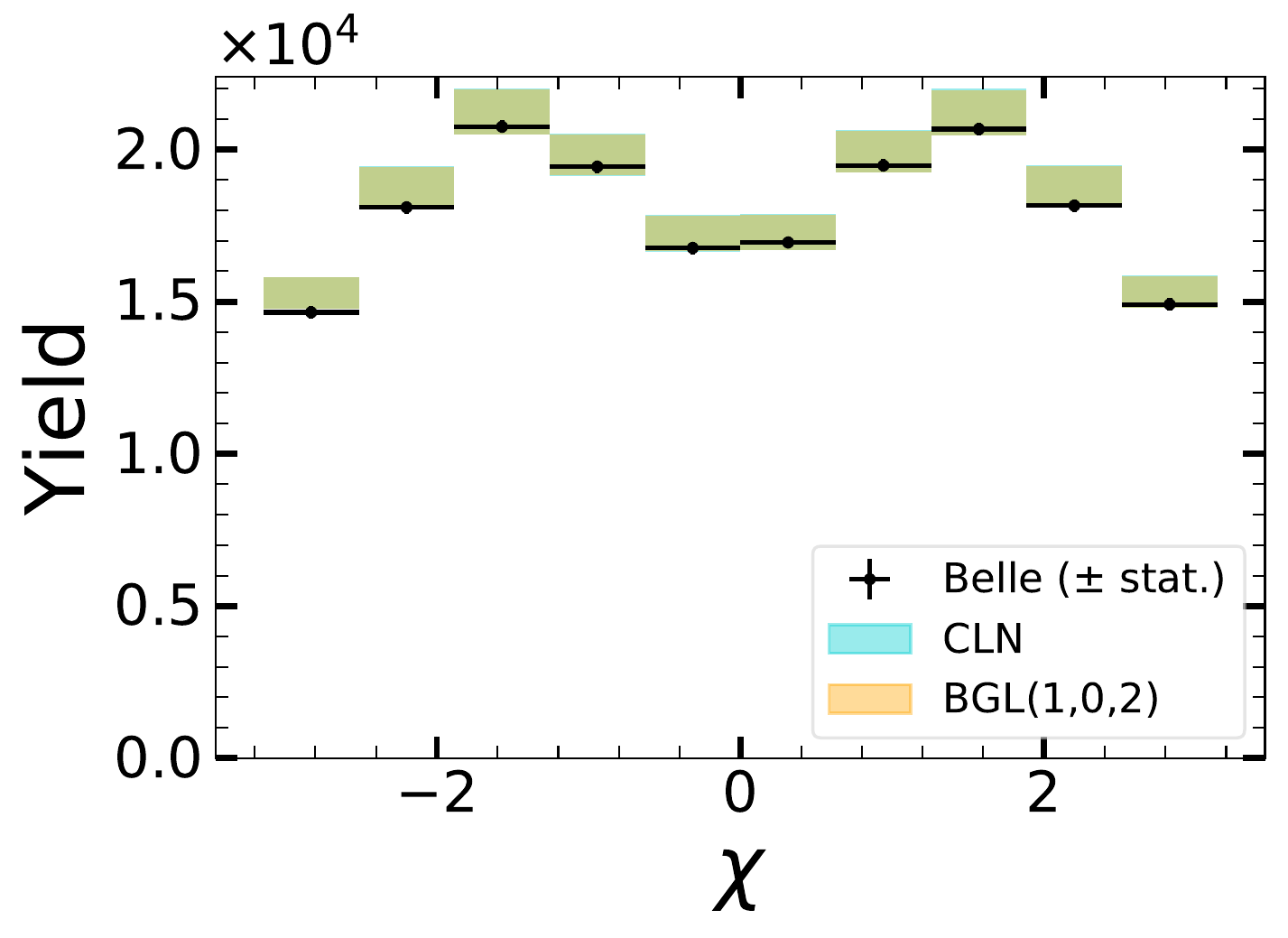}
    \end{minipage}
    \caption{The measured binned yields (data points) with statistical uncertainties for each observable of the $B^{0} \to D^{*-} \ell^{+} \nu_{\ell}$ decay overlaid with both the CLN (cyan) and BGL(1,0,2) (orange) parametrization fit results. The statistical and systematic uncertainties in the fits are determined via the toy MC method with the Cholesky decomposition and added in quadrature for these graphs. As the analysis uses a forward-folding approach, the yields predicted by the fit results are also binned. The results from the CLN and BGL parametrizations are in agreement with each other and with the data.}
    \label{fig:both_fits}
\end{figure*}

The fit parameter distributions from both the statistical and systematic toy MC samples are found to be Gaussian, where the central value of each parameter is taken from the nominal fit without the toy MC method and the statistical and systematic uncertainties are taken from the standard deviations of the distributions from their respective toy MC samples. The mean values for the parameters agree between the fit with no toy MC and the toy MC distributions from fluctuating the data based on the statistical or systematic uncertainties.
We observe in Fig.~\ref{fig:both_fits} that the yields obtained from the CLN and BGL(1,0,2) parametrizations are in excellent agreement with each other, and model the data well. The results for CLN and BGL(1,0,2) are given in the first columns of Tables~\ref{tab:allCLN} and \ref{tab:BGL_noCons}, respectively, noting that the overall sign of the BGL coefficients is arbitrary. The systematic uncertainties are largest in $\mathcal{F}(1)\eta_{\rm EW}|V_{cb}|$, as it is associated to overall normalization, but the central values are consistent between CLN and BGL(1,0,2) within statistical uncertainty alone. The reported $B^{0} \to D^{*-}\ell^{+} \nu_{\ell}$ branching fractions are found to be in agreement with the previous average \cite{cite-PDG}, as well as the Belle analysis.

The systematic uncertainties reported here differ to results presented in Ref.~\cite{cite-Eiasha}. This is attributable to using a forward-folding method where fits were applied to background-subtracted data and all correlations between uncertainties were taken to be linear. This may not consistently represent the exact nature of the correlations present in the data, which were taken into account in the original study.

\section{Alternative CLN scenarios}
\label{sec:CLNs}
The form factor ratios for $B^{0} \to D^{*-}\ell^+ \nu_\ell$ are defined as ratios of the vector and axial-vector form factors: 
\begin{eqnarray}\label{eq:FF-ratios}
R_{1}(w) &=& \frac{h_{V}(w)}{h_{A_{1}}(w)}, \nonumber\\
R_{2}(w)&=& \frac{h_{A_{3}}(w)+ r h_{A_{2}}(w)}{h_{A_{1}}(w)}.
\end{eqnarray}
In the heavy quark limit, $h_{A_{1}} = h_{A_{3}} = h_{V} = \xi$ and $h_{A_{2}} = 0$, where $\xi$ is the Isgur-Wise function \cite{IsgurWise1, IsgurWise2}.
These form factors can be expanded in powers of $\Lambda_{\rm QCD}/m_{c,b}$ and $\alpha_{s}$. It is convenient to parametrize deviations from the heavy quark limit using Eq.~\ref{eq:FF-ratios}, which satisfy $R_{1,2}(w) = 1 + \mathcal{O}(\Lambda_{\rm QCD}/m_{c,b},\alpha_{s})$ in the $m_{c,b}\gg \Lambda_{\rm QCD}$ limit. 

We introduce the parametrization discussed in Ref.~\cite{Bernlochner:2017xyx}, ``CLNnoR", in which 
\begin{eqnarray}\label{eq:FF-ratios_noR}
R_{1}(w) &=& R_{1}(1)+(w-1) R'_{1}(1), \nonumber\\
R_{2}(w) &=& R_{2}(1)+(w-1) R'_{2}(1),
\end{eqnarray}
and fit $R'_{1}(1)$ and $R'_{2}(1)$ as additional floating parameters. CLNnoR is a simple modification of the CLN parametrization that removes QCD sum rule inputs and the condition $R_{1,2}(w) = 1 + \mathcal{O}(\Lambda_{\rm QCD}/m_{c,b}, \alpha_{s})$ but still relies on heavy quark symmetry and model-dependent input on subleading Isgur-Wise functions due to constraints on the cubic polynomial used to describe the form factor $h_{A_1}$. 
These heavy quark symmetry constraints can be further loosened to form the model ``CLNnoHQS", in which  $h_{A_{1}}$ from Eq.~\ref{eq:cln-eqs} is parametrized by a quadratic polynomial in $z$, with unconstrained coefficients:
\begin{eqnarray}
h_{A_{1}} (w)&=&  h_{A_{1}}(1)[1-8\rho^{2}z + (53c_{D^{*}} - 15)z^{2}],
\end{eqnarray}
where setting $c_{D^{*}}$ = $\rho^{2}$ will return the original CLN parametrization truncated to its second order.

\begin{figure*}
    \centering
    \begin{minipage}{0.48\linewidth}
        \includegraphics[width=0.98\linewidth]{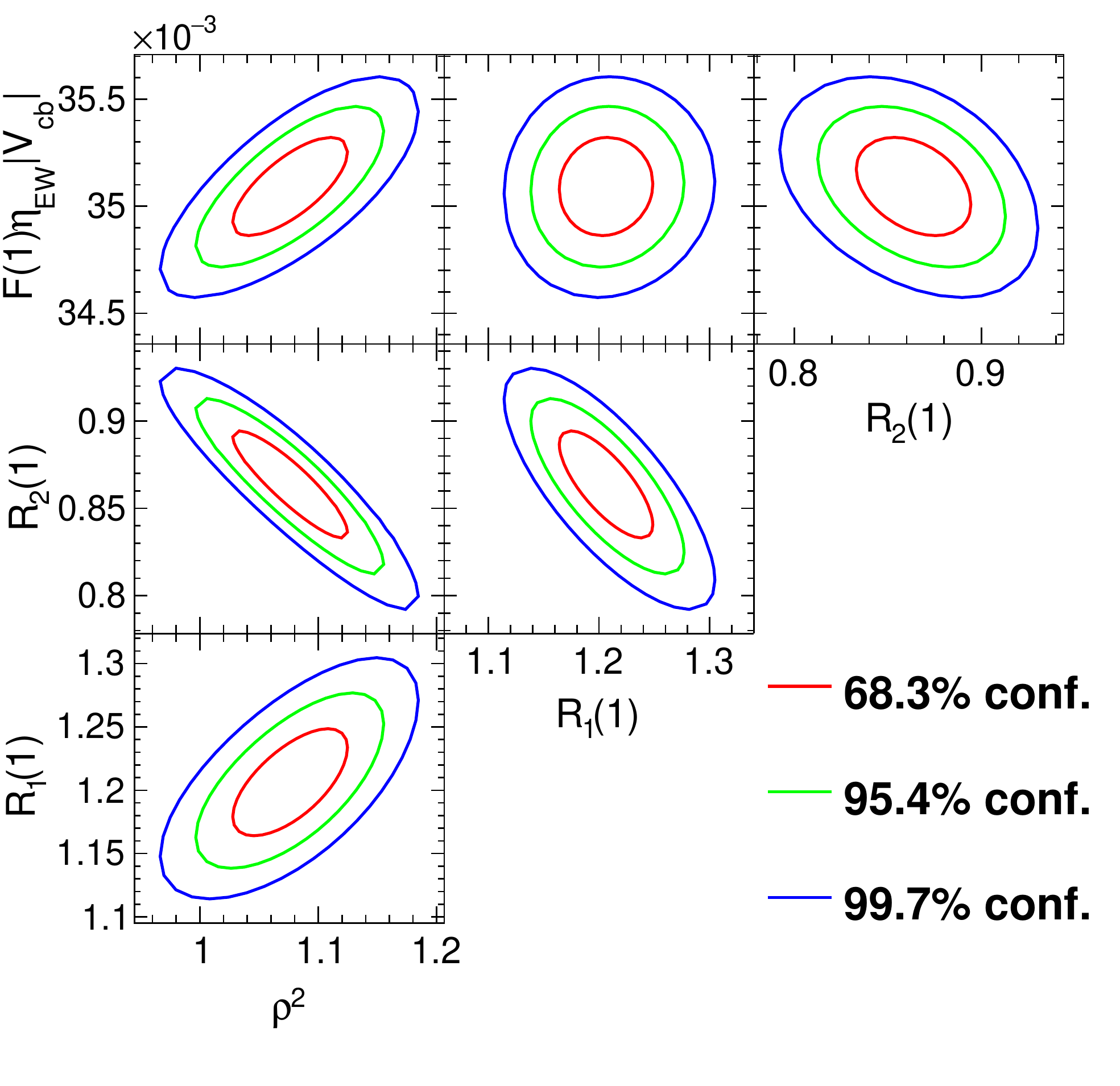}
    \end{minipage}
    \begin{minipage}{0.49\linewidth}
        \includegraphics[width=\linewidth]{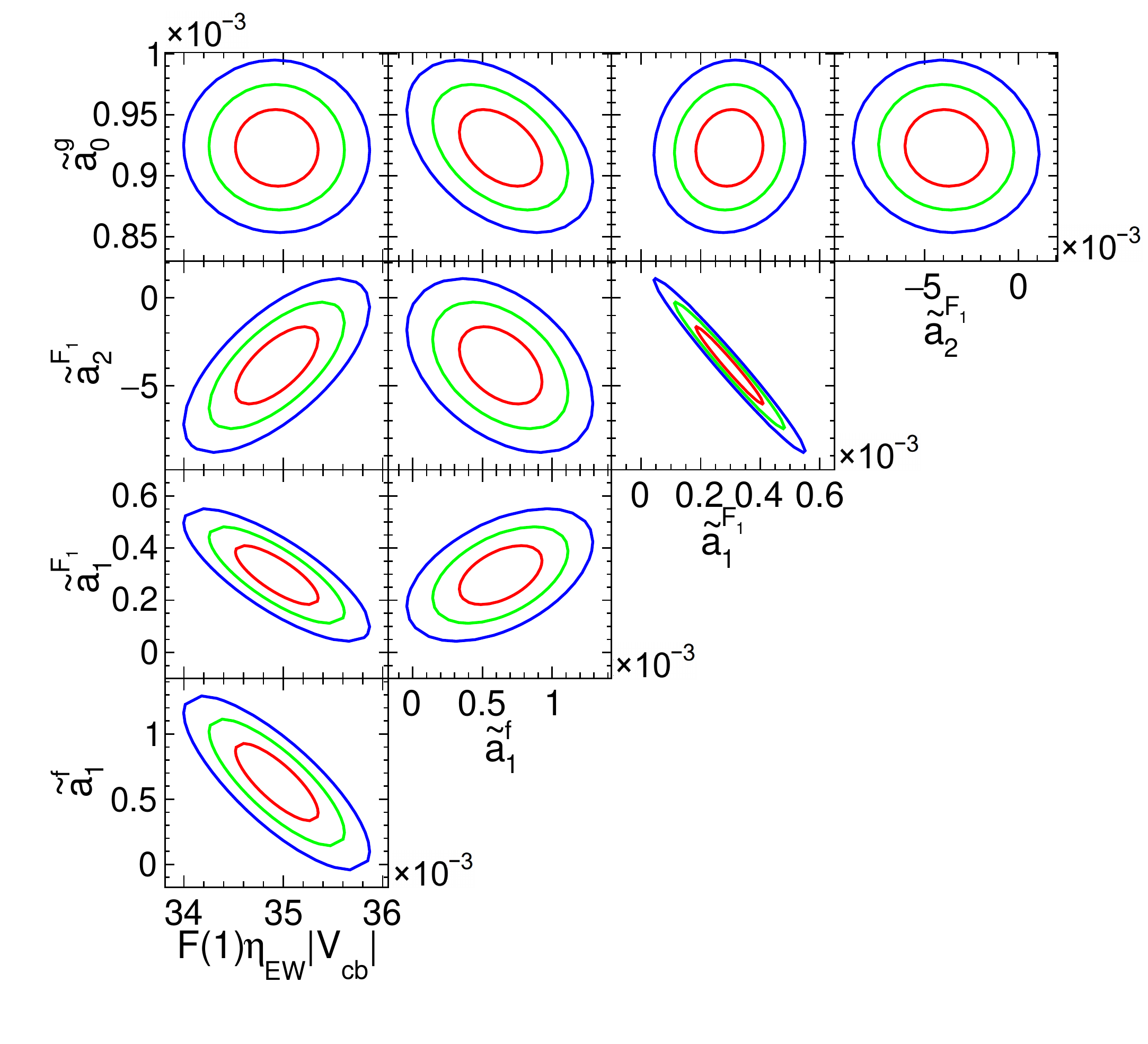}
    \end{minipage}
    \caption{2-D contours of pairs of fit parameters for CLN (left) and BGL(1,0,2) (right) at $68.3\%$, $95.4\%$ and $99.7\%$ confidence levels. Note that from Eq.~\ref{eq:a0f_to_Vcb}, the correlations for $\tilde{a}_0^f$ and $\mathcal{F}(1)\eta_{\rm EW}|V_{cb}|$ are equal in BGL and therefore only the latter is shown.}
    \label{fig:contour_CLN_OG}
\end{figure*}

We performed fits to the $B^{0} \to D^{*-}\ell^+ \nu_\ell$ data for the three different CLN scenarios and the results are summarized in Table \ref{tab:allCLN}. Large uncertainties are seen in the results for CLNnoR and CLNnoHQS.
The contours for the standard CLN and BGL(1,0,2) models are shown in Fig.~\ref{fig:contour_CLN_OG}, where smooth curves indicate that the true minimum in the log-likelihood function was found and therefore the variation in parameters with respect to each other around this minimum can be determined.

\begin{table*}[t]
\centering
\caption{Fitted parameters for the CLN, CLNnoR and CLNnoHQS scenarios. The uncertainties listed are statistical and systematic, respectively. The branching ratios are obtained from the fit. No additional input from LQCD has been used.}
\renewcommand{\arraystretch}{1.3}
\begin{tabularx}{0.8\linewidth}{YYYY}
\hline \hline
Parameter & CLN & CLNnoR & CLNnoHQS\\ \hline
$\rho^{2}$ & $1.09 \pm 0.04 \pm 0.05$ & $0.89 \pm 0.09 \pm 0.15$ & $0.93 \pm 0.38 \pm 0.39$ \\
$R_{1}(1)$ & $1.20 \pm 0.03 \pm 0.02$ & $2.01 \pm 0.43 \pm 0.46$ & $2.10 \pm 0.61 \pm 0.53$ \\
$R_{2}(1)$ & $0.86 \pm 0.02 \pm 0.01$ & $0.83 \pm 0.05 \pm 0.04$ & $0.77 \pm 0.42 \pm 0.38$ \\
$R'_{1}(1)$ & $-0.12$ (fixed) & $-3.50 \pm 1.77 \pm 1.85$ & $-3.86 \pm 2.50 \pm 2.15$ \\
$R'_{2}(1)$ & $0.11$ (fixed) & $0.28 \pm 0.14 \pm 0.16$ & $0.50 \pm 1.45 \pm 1.32$ \\
$c_{D}*$ & $\rho^{2}$ (fixed) & $\rho^{2}$ (fixed) & $1.00 \pm 1.73 \pm 1.59$ \\
$\mathcal{F}(1)\eta_{\rm EW}|V_{cb}|\times10^{3}$ & $35.63 \pm 0.18 \pm 0.80$ & $34.29 \pm 0.67 \pm 1.33$ & $34.23 \pm 0.63 \pm 1.30$ \\
\hline
$\mathcal{B}(B^{0} \to D^{*-}\ell^+ \nu_\ell)$ & $5.04$ & $5.04$ & $5.04$\\
$\chi^{2}$/ndf & $40.8/36$ & $34.8/34$ & $34.7/33$\\
\hline \hline
\end{tabularx}
\label{tab:allCLN}
\end{table*}
\begin{figure*}
 \centering
\includegraphics[width=0.4\linewidth]{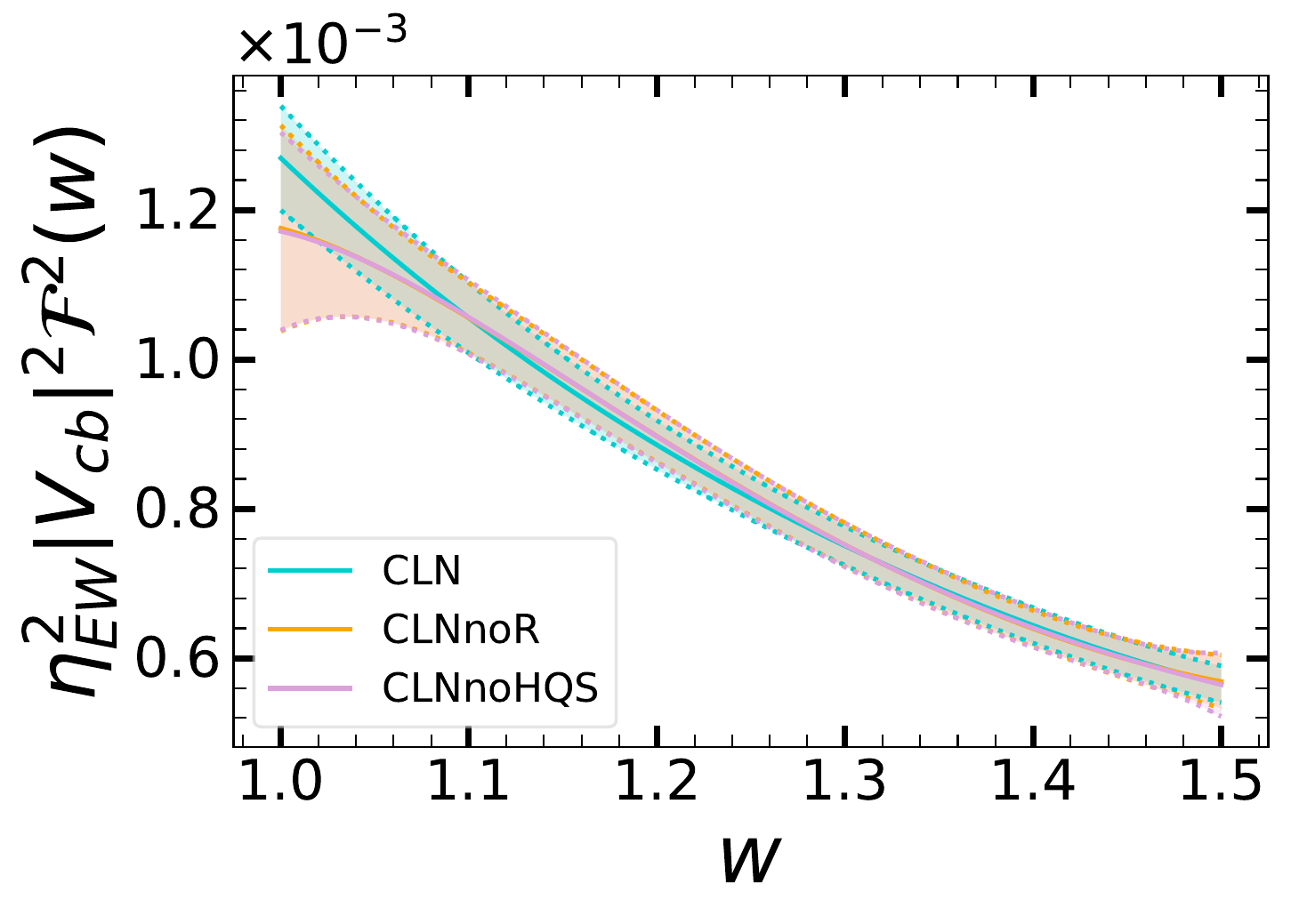}
\includegraphics[width=0.4\linewidth]{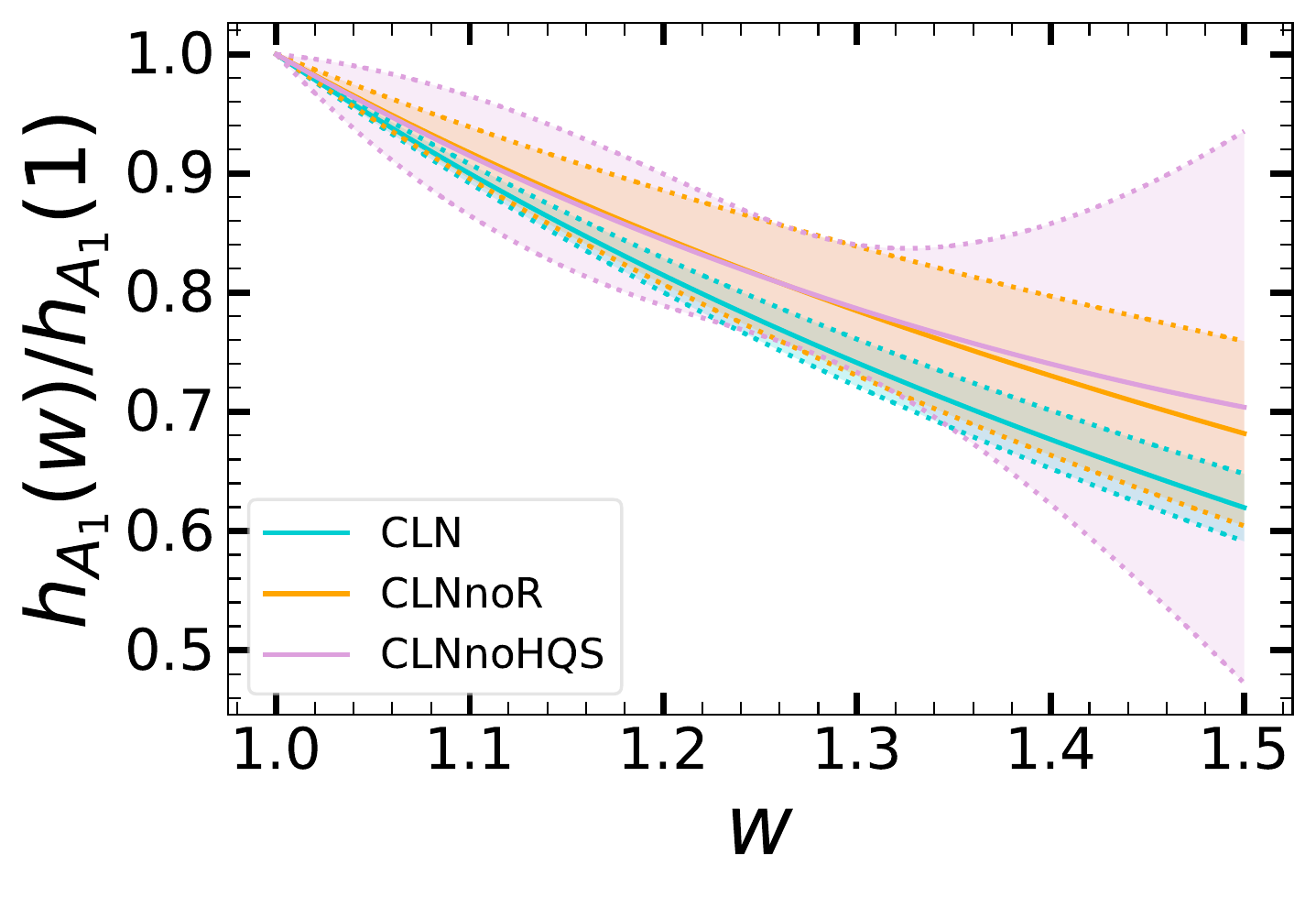}
\includegraphics[width=0.4\linewidth]{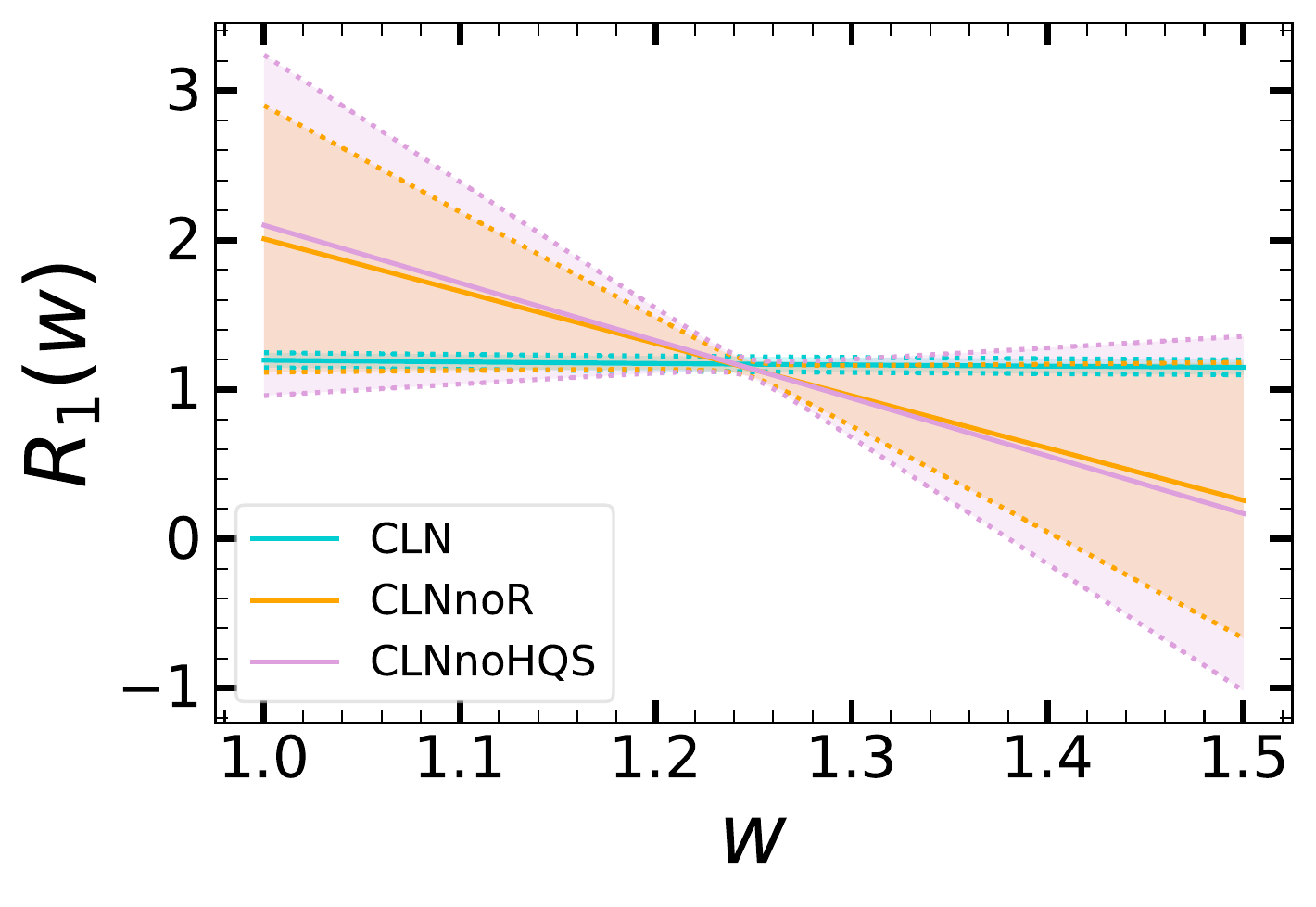}
\includegraphics[width=0.4\linewidth]{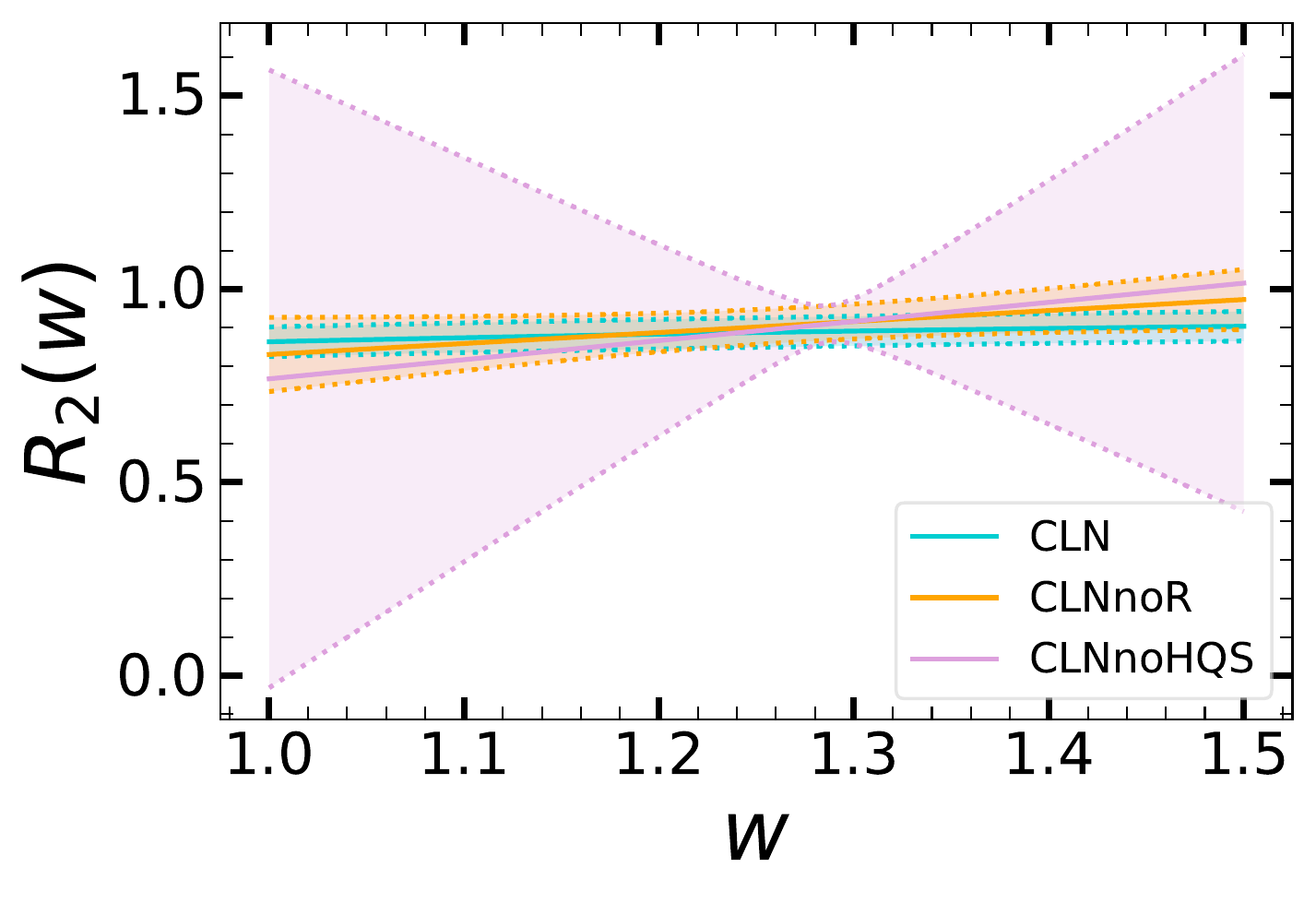}
\caption{$\eta_{\rm EW}^2|V_{cb}|^2\mathcal{F}^2$, $h_{A_1}(w)/h_{A_1}(1)$, $R_1$ and $R_2$ as a function of the hadronic recoil, $w$, for the CLN, CLNnoR and CLNnoHQS scenarios. The solid lines correspond to the central values from the fit with no toy MC and the uncertainty bands correspond to the combination of the statistical and systematic uncertainties obtained from the standard deviation of the distribution of the function values at each point in $w$.}
 \label{fig:clnfit_noLQCD}
 \end{figure*}
Figure~\ref{fig:clnfit_noLQCD} shows the form factor ratios $R_{1}$ and $R_{2}$, the form factor $h_{A_{1}}$ divided by its value at $w=1$, and  $\eta_{\rm EW}^2|V_{cb}|^2\mathcal{F}^2$ for the three CLN fit configurations as a function of the hadronic recoil, $w$. The combination of the statistical and systematic uncertainties in the fits are shown for each configuration, obtained from the toy MC method with the Cholesky decomposition. We note that in both CLN and BGL, the uncertainty in the plots of $h_{A_1}(w)$ reduces to zero at $w=1$, and has been presented as a ratio to avoid assuming the value and uncertainty of $h_{A_1}(1)$. Including this value and its uncertainty would have no effect on the overall convergence of the fits and is therefore beyond the scope of this analysis.

The distribution of $\eta_{\rm EW}^2|V_{cb}|^2\mathcal{F}^2$ shows that the CLNnoR and CLNnoHQS scenarios have large uncertainty at low hadronic recoil, with their central values diverging from the standard CLN value, but remaining compatible within uncertainty. At larger values of hadronic recoil, this form factor becomes less sensitive to the fit parameters. The results of CLNnoR and CLNnoHQS for the form factor ratios $R_1(w)$ and $R_2(w)$ show large uncertainties near zero and maximal recoil and remain consistent with CLN. The compatibility between these two configurations and the more model-dependent CLN are an indication that the heavy quark symmetry assumed in the structure of CLN remains present when these assumptions are no longer included. This is in contrast to the study performed in Ref.~\cite{Bernlochner:2017xyx} using a preliminary tagged Belle dataset \cite{Abdesselam:2017kjf}. Our results are found to be more similar to those from a study based on BGL~\cite{GAMBINO2019386}, where the plots of form factor ratios are consistent with expected theoretical results for heavy quark symmetry.

\section{Higher order BGL}
\label{sec:BGLs}
\begin{figure*}[t]
 \centering
\includegraphics[width=0.4\linewidth]{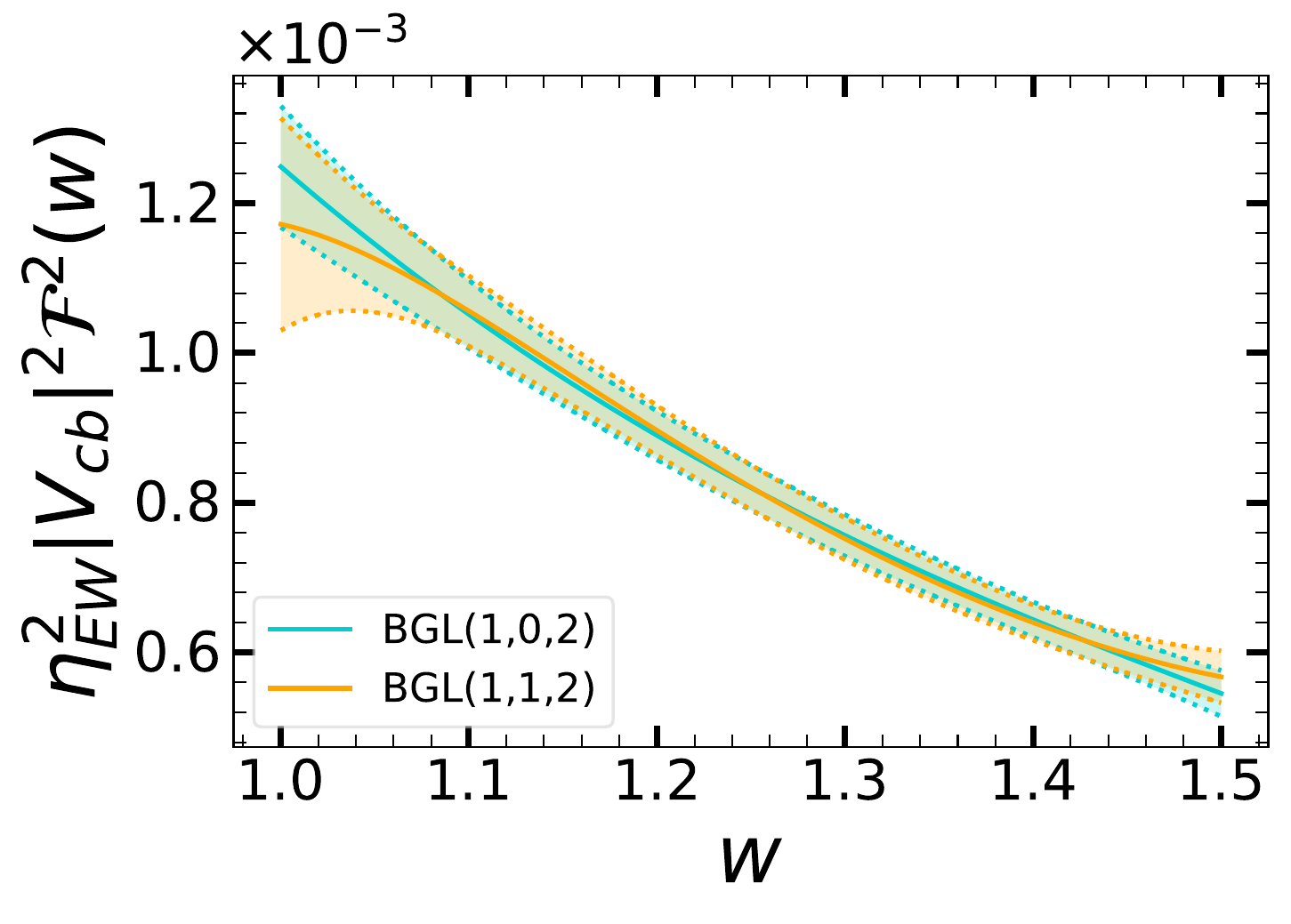}
\includegraphics[width=0.4\linewidth]{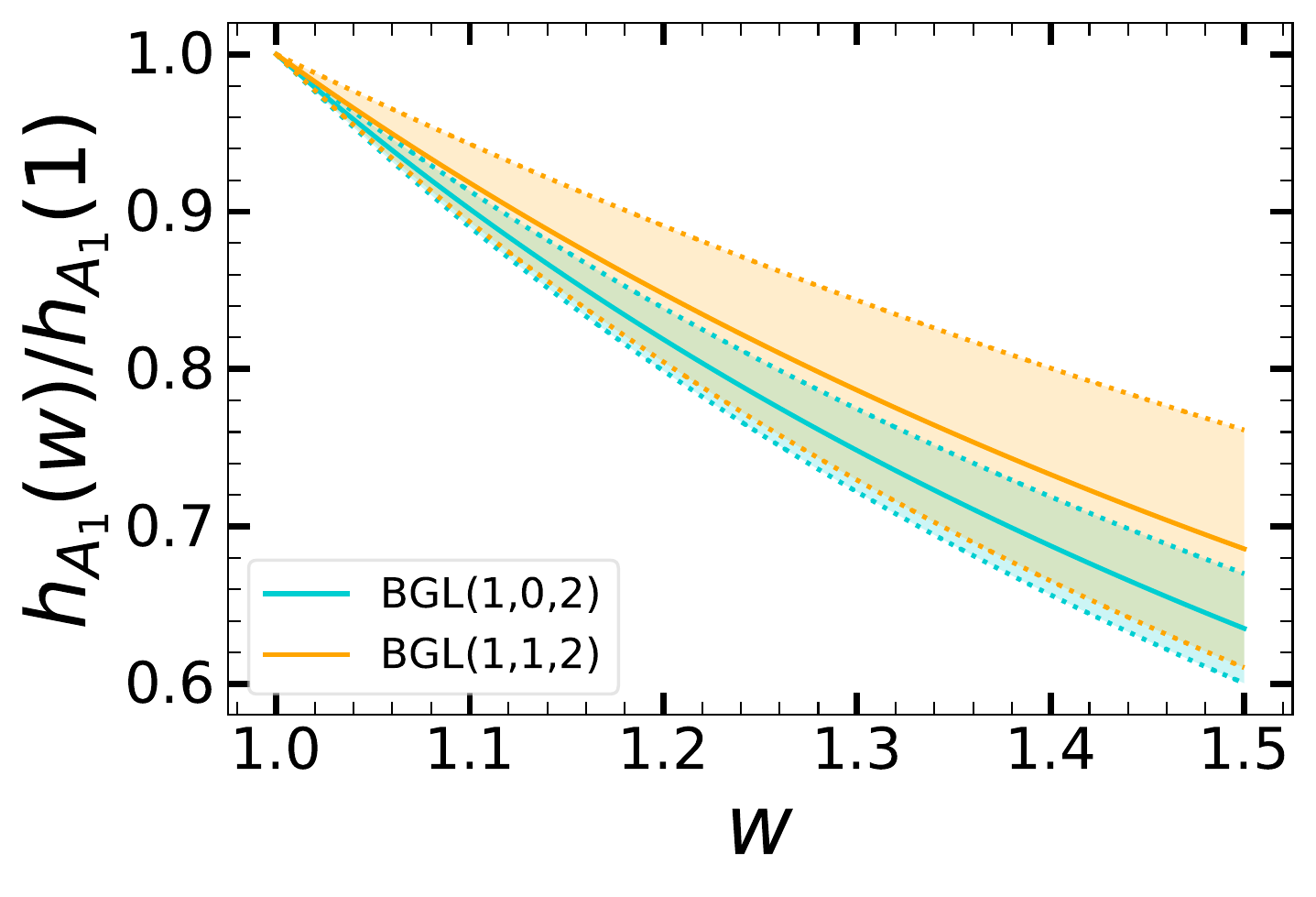}
\includegraphics[width=0.4\linewidth]{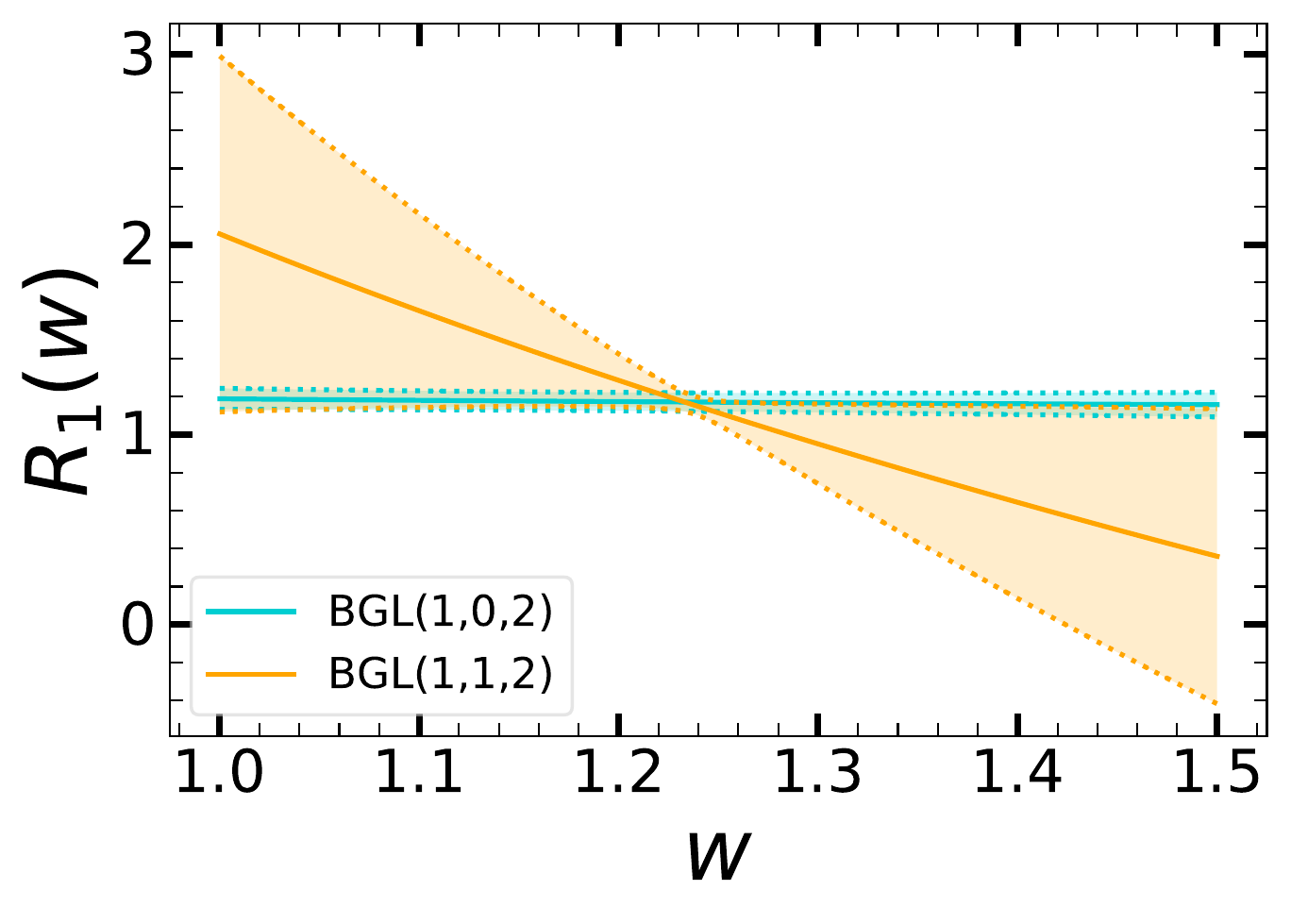}
\includegraphics[width=0.4\linewidth]{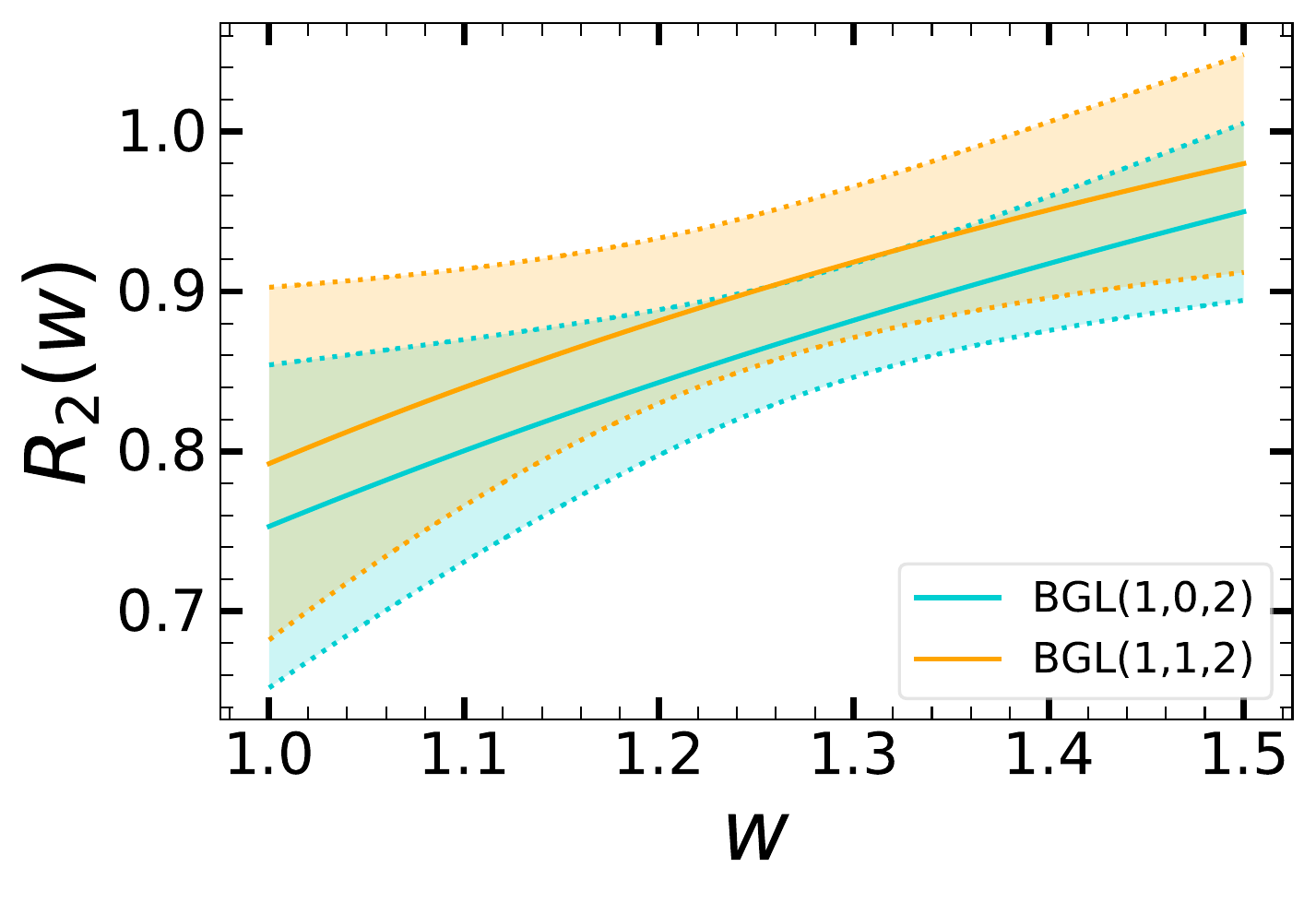}
\caption{The form factors and ratios $\eta_{\rm EW}^2|V_{cb}|^2\mathcal{F}^2$, $h_{A_1}(w)/h_{A_1}(1)$, $R_1$ and $R_2$ as a function of the hadronic recoil, $w$, for BGL(1,0,2) and BGL(1,1,2). The central values and uncertainty bands are calculated with the same method as Fig.~\ref{fig:clnfit_noLQCD}.}
 \label{fig:bglfit_noLQCD}
 \end{figure*}

The form factors, $f$, $g$ and $\mathcal{F}_1$ in the BGL parametrization are defined as a power series of $z(w)$. We can perform fits to the data for truncations at varying order to search for a plateau in fit results at increasing order and therefore reduce the model dependence.
The results of the BGL fit with expansions $(1,0,2)$ and $(1,1,2)$ are given in Table \ref{tab:BGL_noCons}, where fits to expansions with more free parameters do not converge with the given dataset using MINUIT. From Table \ref{tab:BGL_noCons}, we note that increasing the power in the $z$ expansions also has an impact on the results, including shifting the value obtained for $|V_{cb}|$ lower by $3\%$ but remaining compatible within systematic uncertainties. This shift can also be seen in Fig.~\ref{fig:bglfit_noLQCD}, where the introduction of additional parameters to the fit leads to large deviations and large uncertainties in form factors, similar to the effects seen in CLNnoR and CLNnoHQS. These shapes are consistent with the results seen in Ref.~\cite{GAMBINO2019386} where uncertainties dominate the hadronic recoil end points. Form factors obtained by LQCD calculations \cite{Kaneko:2019vkx} appear to have a shape more consistent with those from the BGL(1,0,2) and CLN configurations. From this, we conclude that adding more free parameters into the fits to this dataset without further constraints introduces instability in both parametrization models and although most results remain consistent within large uncertainties, useful results become difficult to extract.\\

\begin{table*}
\centering
\caption{Fitted parameter values for the BGL(1,0,2) and BGL(1,1,2) configurations. The uncertainties listed are statistical and systematic, respectively. The branching ratios are obtained from the fit, and $\mathcal{F}(1)\eta_{\rm EW}|V_{cb}|$ is calculated from Eq. \ref{eq:a0f_to_Vcb}. Configurations with higher order coefficients do not converge without additional input from LQCD and are therefore not shown.
}
 \renewcommand{\arraystretch}{1.3}
\begin{tabularx}{0.8\linewidth}{YYY}
\hline \hline
Parameter $\times 10^{3}$ & BGL(1,0,2) & BGL(1,1,2)\\ \hline
$\tilde{a}_{0}^{f}$ & $0.512 \pm 0.004 \pm 0.013$ & $0.496 \pm 0.010 \pm 0.020$ \\
$\tilde{a}_{1}^{f}$ & $0.64 \pm 0.19 \pm 0.33$ & $1.38 \pm 0.40 \pm 0.67$ \\
$\tilde{a}_{2}^{f}$ & $0.0$ (fixed) & $0.0$ (fixed) \\
$\tilde{a}_{0}^{g}$ & $0.93 \pm 0.02 \pm 0.01$ & $1.56 \pm 0.31 \pm 0.33$ \\
$\tilde{a}_{1}^{g}$ & $0.0$ (fixed) & $-22.63 \pm 11.03 \pm 11.62$ \\
$\tilde{a}_{2}^{g}$ & $0.0$ (fixed) & $0.0$ (fixed) \\
$\tilde{a}_{1}^{\mathcal{F}_1}$ & $0.30 \pm 0.07 \pm 0.09$ & $0.38 \pm 0.10 \pm 0.14$ \\
$\tilde{a}_{2}^{\mathcal{F}_1}$ & $-3.88 \pm 1.46 \pm 1.34$ & $-3.75 \pm 1.53 \pm 1.59$ \\
$\mathcal{F}(1)\eta_{\rm EW}|V_{cb}|$ & $35.34 \pm 0.27 \pm 0.87$ & $34.23 \pm 0.66 \pm 1.40$ \\
\hline
$\mathcal{B}(B^{0} \to D^{*-}\ell^+ \nu_\ell)$ & $5.04$ & $5.04$\\
$\chi^{2}$/ndf & $38.6/35$ & $34.8/34$\\
\hline \hline
\end{tabularx}
\label{tab:BGL_noCons}
\end{table*}

\section{Additional data from LQCD}
\label{sec:LQCD}
The use of data from LQCD calculations of form factors at non-zero recoil is an important recent development in the measurement of $|V_{cb}|$~\cite{Aviles-Casco:2017nge}. By providing precise input near zero recoil, form factors can be much better constrained than from experimental information alone, owing to the presence of a low-momentum (slow) pions affecting efficiencies in this region of phase space. Results have been presented in Ref.~\cite{Vaquero:2019ary} (Fermilab/MILC collaboration), which has a blinded normalization factor, and Ref.~\cite{Kaneko:2019vkx} (JLQCD collaboration), which presents preliminary unblinded results.

This paper examines the impact of LQCD inputs on constraining higher order parametrizations (BGL) and the impact of removing theoretical constraints (CLN), both leading to reductions in model dependence in the evaluation of $|V_{cb}|$. Although $\mathcal{F}(1)\eta_{\rm EW}|V_{cb}|$ values are quoted, the results are based on preliminary LQCD inputs that are subject to change. While other studies have used form factors evaluated at non-zero recoil  in discussions on implications for $R(D^{*}) $\cite{Jaiswal:2020wer} or have included additional form factor constraints from light cone sum rules~\cite{Jaiswal:2017rve}, this analysis features a novel approach for including data from LQCD as additional constraints for determining $B^{0} \to D^{*-}\ell^+ \nu_\ell$ form factors in various parametrization scenarios. Four points related to $h_{V}$ and $h_{A_1}$ at different values of hadronic recoil were used:
\begin{align}
     h_{A_1}(1.04)/h_{A_1}(1) &= 0.9534 \cdot  (1 \pm 1.3\% ~^{+3.4\%}_{-1.8\%}), \nonumber\\
     h_{A_1}(1.08)/h_{A_1}(1) &=  0.9093  \cdot (1\pm 1.3\% ~^{+3.5\%}_{-1.8\%}), \nonumber\\
     h_{V}(1.04)/h_{V}(1) &= 0.9403 \cdot(1 \pm 2.4\% ~^{+6.8\%}_{-3.7\%}), \nonumber\\
     h_{V}(1.08)/h_{V}(1) &= 0.9040  \cdot (1 \pm 2.4\% ~^{+6.9\%}_{-3.7\%}),
     \label{eq:LQCD_points_OG}
\end{align}
with $h_{A_1}(1) = 0.906$ and $h_{V}(1) = 1.18$ \cite{Kaneko:2019vkx}. The systematic uncertainties have been symmetrized by conservatively taking the larger of the upper and lower errors that were provided.
These points were appended to the $N_{\rm obs}$ vector from Eq.~\ref{eq:chi2}, expanding it to size $40 + 4$, while the correlations between these values (obtained from preliminary estimates via private correspondence with the JLQCD group and listed in Table~\ref{tab:LQCD_incorr}) were appended to the original statistical covariance matrix in a block-diagonal manner. The $N_{\rm exp}$ vector was expanded to include the calculated values for the corresponding form factors given a set of model parameters.

\subsection{Impact of LQCD constraints on fits}
The $\chi^2$ minimization approach was applied with the Cholesky toy MC method to obtain results for various BGL and CLN parametrizations and are presented in presented in Tables~\ref{tab:allCLN_LQCD} and \ref{tab:BGL_LQCD}.

\begin{table*}
\centering
\caption{Fitted parameters for the CLN, CLNnoR and CLNnoHQS scenarios using data from LQCD calculations of form factors at non-zero recoil as additional constraints. The uncertainties listed are statistical and systematic, respectively. The branching ratio is obtained from the fit.
}
\renewcommand{\arraystretch}{1.3}
\begin{tabularx}{0.8\linewidth}{YYYY}
\hline \hline
Parameter & CLN & CLNnoR & CLNnoHQS\\ \hline
$\rho^{2}$ & $1.10 \pm 0.04 \pm 0.05$ & $1.05 \pm 0.04 \pm 0.06$ & $0.80 \pm 0.18 \pm 0.23$ \\
$R_{1}(1)$ & $1.21 \pm 0.03 \pm 0.02$ & $1.29 \pm 0.02 \pm 0.02$ & $1.25 \pm 0.02 \pm 0.02$ \\
$R_{2}(1)$ & $0.86 \pm 0.02 \pm 0.01$ & $0.80 \pm 0.05 \pm 0.04$ & $0.99 \pm 0.17 \pm 0.16$ \\
$R'_{1}(1)$ & $-0.12$ (fixed) & $-0.45 \pm 0.10 \pm 0.05$ & $-0.26 \pm 0.11 \pm 0.10$ \\
$R'_{2}(1)$ & $0.11$ (fixed) & $0.26 \pm 0.13 \pm 0.13$ & $-0.39 \pm 0.58 \pm 0.52$ \\
$c_{D}*$ & $\rho^{2}$ (fixed) & $\rho^{2}$ (fixed) & $-0.06 \pm 0.71 \pm 0.77$ \\
$\mathcal{F}(1)\eta_{\rm EW}|V_{cb}|\times10^{3}$ & $35.63 \pm 0.19 \pm 0.76$ & $35.29 \pm 0.23 \pm 0.85$ & $34.96 \pm 0.32 \pm 0.96$ \\
\hline
$\mathcal{B}(B^{0} \to D^{*-}\ell^+ \nu_\ell)$ & $5.04$ & $5.04$ & $5.04$\\
$\chi^{2}$/ndf & $42.3/40$ & $38.8/38$ & $37.5/37$\\
\hline \hline
\end{tabularx}
\label{tab:allCLN_LQCD}
\end{table*}

\begin{table*}
\centering
\caption{Fitted parameters for the BGL(1,0,2), BGL(1,1,2) and BGL(2,2,2) configurations using data from LQCD calculations of form factors at non-zero recoil as additional constraints. The uncertainties listed are statistical and systematic, respectively. The branching ratio is obtained from the fit and $\mathcal{F}(1)\eta_{\rm EW}|V_{cb}|$ is calculated from Eq.~\ref{eq:a0f_to_Vcb}}.

 \renewcommand{\arraystretch}{1.3}
\begin{tabularx}{0.8\linewidth}{YYYY}
\hline \hline
Parameter $\times 10^{3}$ & BGL(1,0,2) & BGL(1,1,2) & BGL(2,2,2)\\ \hline
$\tilde{a}_{0}^{f}$ & $0.512 \pm 0.004 \pm 0.013$ & $0.511 \pm 0.004 \pm 0.013$ & $0.507 \pm 0.004 \pm 0.013$ \\
$\tilde{a}_{1}^{f}$ & $0.62 \pm 0.18 \pm 0.30$ & $0.67 \pm 0.17 \pm 0.30$ & $1.43 \pm 0.64 \pm 0.84$ \\
$\tilde{a}_{2}^{f}$ & $0.0$ (fixed) & $0.0$ & $-19.81 \pm 17.52 \pm 18.50$ \\
$\tilde{a}_{0}^{g}$ & $0.94 \pm 0.02 \pm 0.01$ & $1.00 \pm 0.02 \pm 0.02$ & $0.98 \pm 0.02 \pm 0.05$ \\
$\tilde{a}_{1}^{g}$ & $0.0$ (fixed) & $-2.35 \pm 0.61 \pm 0.66$ & $0.01 \pm 0.97 \pm 1.74$ \\
$\tilde{a}_{2}^{g}$ & $0.0$ (fixed) & $0.0$ (fixed) & $-38.62 \pm 1.19 \pm 3.81$ \\
$\tilde{a}_{1}^{\mathcal{F}_1}$ & $0.31 \pm 0.06 \pm 0.08$ & $0.30 \pm 0.06 \pm 0.08$ & $0.29 \pm 0.07 \pm 0.08$ \\
$\tilde{a}_{2}^{\mathcal{F}_1}$ & $-4.01 \pm 1.13 \pm 1.01$ & $-3.68 \pm 1.26 \pm 1.20$ & $-3.06 \pm 1.48 \pm 1.24$ \\
$\mathcal{F}(1)\eta_{\rm EW}|V_{cb}|$ & $35.32 \pm 0.24 \pm 0.87$ & $35.28 \pm 0.24 \pm 0.87$ & $35.02 \pm 0.29 \pm 0.88$ \\
\hline
$\mathcal{B}(B^{0} \to D^{*-}\ell^+ \nu_\ell)$ & $5.04$ & $5.04$ & $5.04$\\
$\chi^{2}$/ndf & $40.3/39$ & $38.9/38$ & $37.7/36$\\
\hline \hline
\end{tabularx}
\label{tab:BGL_LQCD}
\end{table*}

Most fit parameters are reasonably consistent between the different configurations, while the values of $\mathcal{F}(1)\eta_{\rm EW}|V_{cb}|$ are in agreement in all scenarios. Overall we find that all fit scenarios model the data well, where the CLN configuration has a p-value of $0.37$ and all other scenarios have similar values. We find that in the BGL(2,2,2) scenario the value for $\tilde{a}_{2}^{g}$ will often reach the boundary condition set by Eq.~\ref{eq:eqs-constraints} and so its uncertainties cannot be considered Gaussian-shaped. Instead, they are calculated in an asymmetric manner where the mean of the difference between each value of $\tilde{a}_{2}^{g}$ and its lower bound is taken as the lower uncertainty and the upper uncertainty is taken as the standard deviation minus this number. The measurement and uncertainty of $\mathcal{F}(1)\eta_{\rm EW}|V_{cb}|$ remains unaffected.

\begin{figure*}
     \centering
     \includegraphics[width=0.5\linewidth]{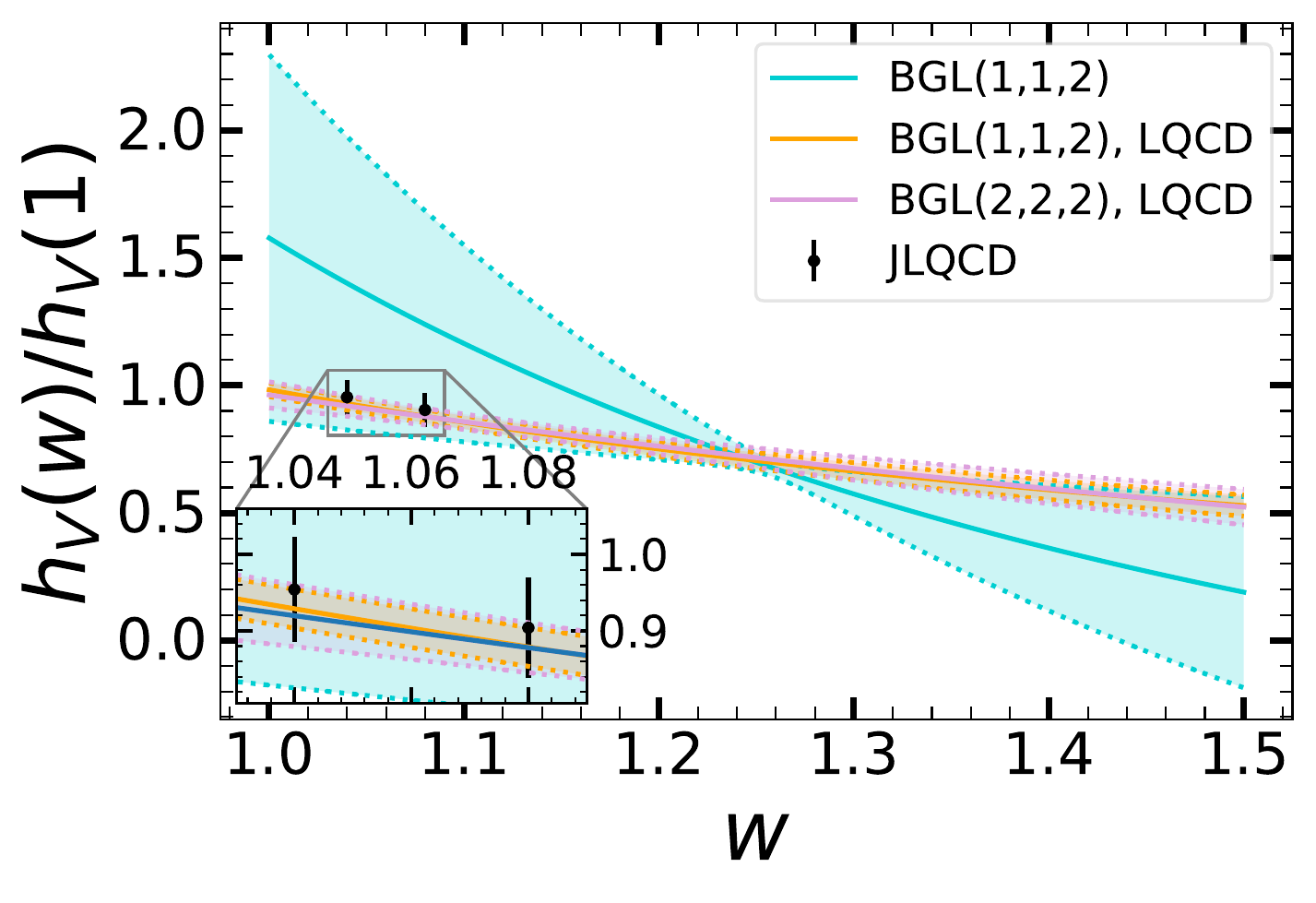}
     \caption{The normalized form factor $h_V(w)/h_V(1)$ as a function of the hadronic recoil, $w$, for BGL(1,1,2) with (orange) and without (cyan) LQCD constraints, and BGL(2,2,2) with LQCD constraints (pink). The zoomed window within the plot is used to emphasize the region around the LQCD inputs for $h_V$. The central values and uncertainty bands are calculated with the same method as Fig.~\ref{fig:clnfit_noLQCD}.}
     \label{fig:LQCD_hV_eg}
 \end{figure*}

\begin{figure*}
 \centering
\includegraphics[width=0.4\linewidth]{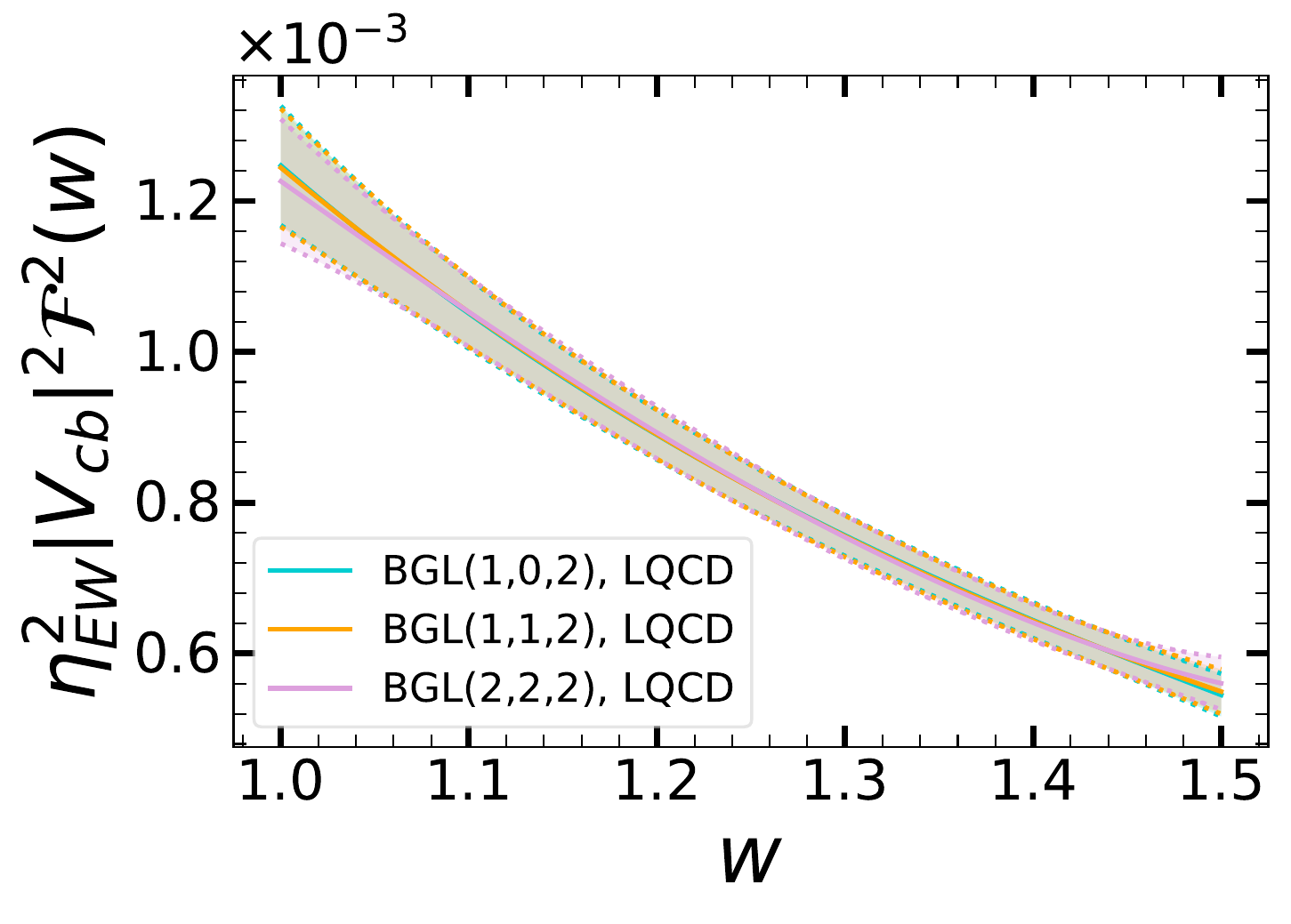}
\includegraphics[width=0.4\linewidth]{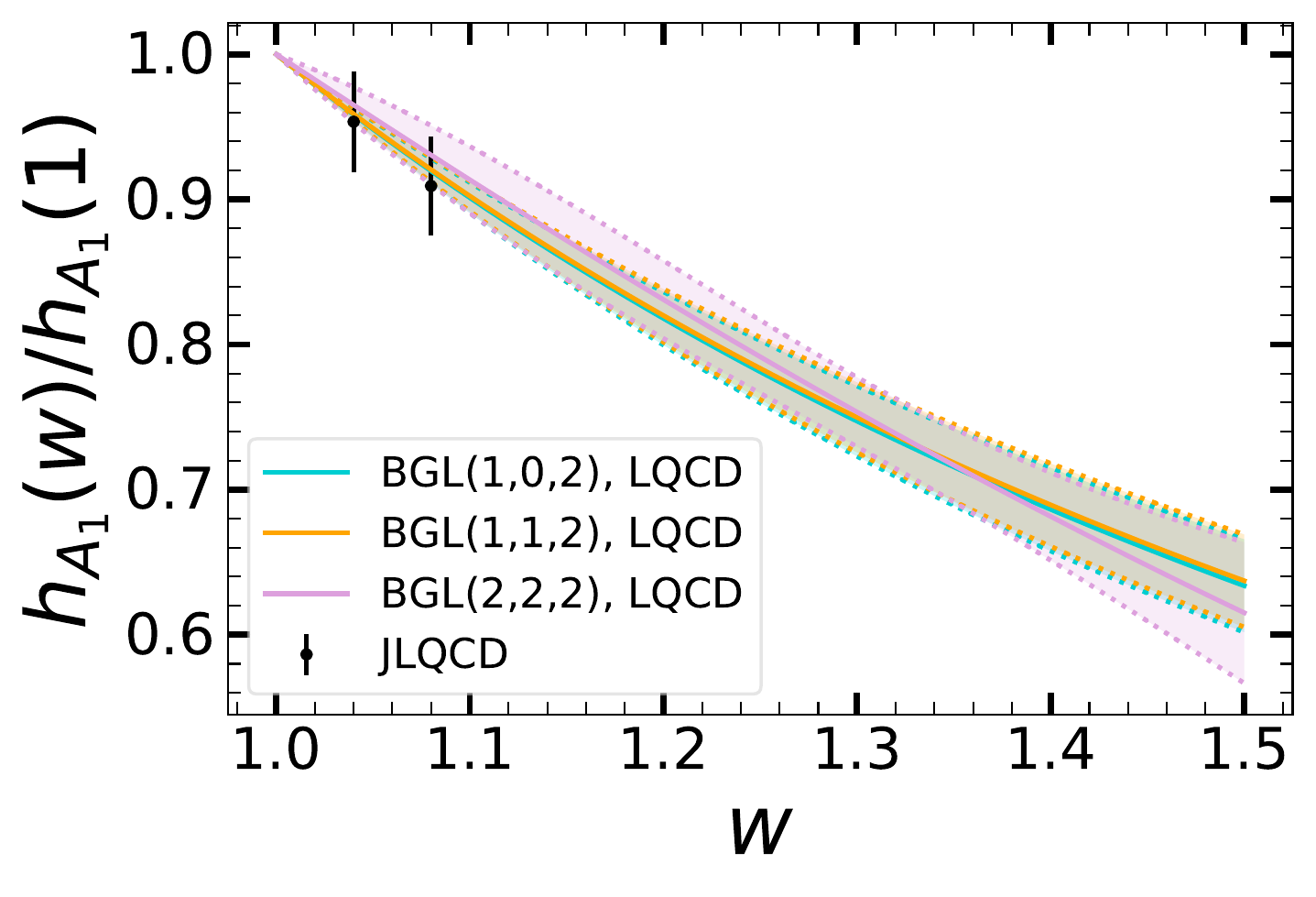}
\includegraphics[width=0.4\linewidth]{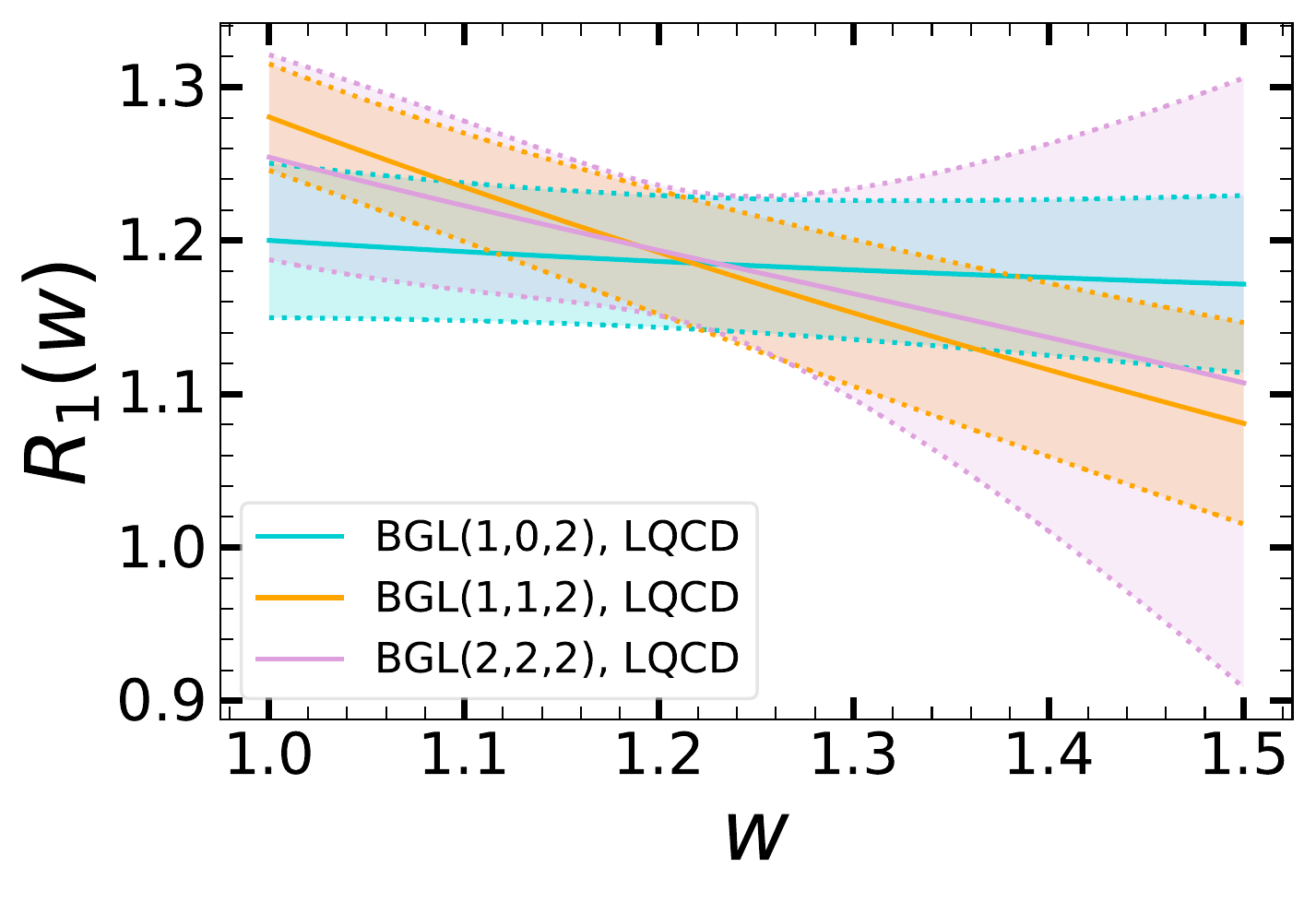}
\includegraphics[width=0.4\linewidth]{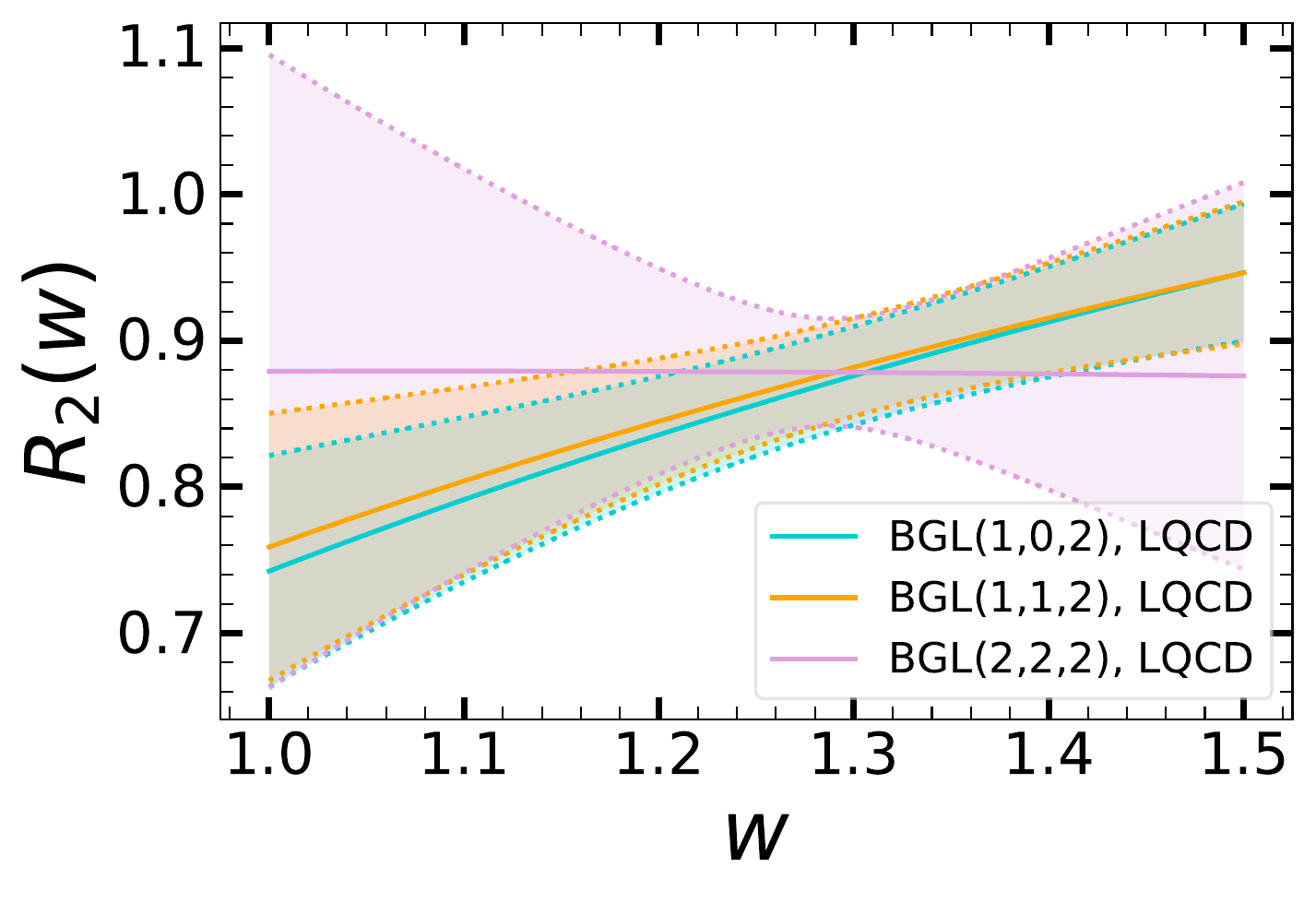}
\caption{The form factors and ratios $\eta_{\rm EW}^2|V_{cb}|^2\mathcal{F}^2$, $h_{A_1}(w)/h_{A_1}(1)$, $R_1$ and $R_2$ as a function of the hadronic recoil, $w$, for BGL(1,0,2), BGL(1,1,2) and BGL(2,2,2) using additional LQCD constraints. The LQCD input has been overlaid in the plot for $h_{A_1}(w)/h_{A_1}(1)$. The central values and uncertainty bands are calculated with the same method as Fig.~\ref{fig:clnfit_noLQCD}.
}
 \label{fig:bglfit_LQCD}
 \end{figure*}

\begin{figure*}
 \centering
\includegraphics[width=0.4\linewidth]{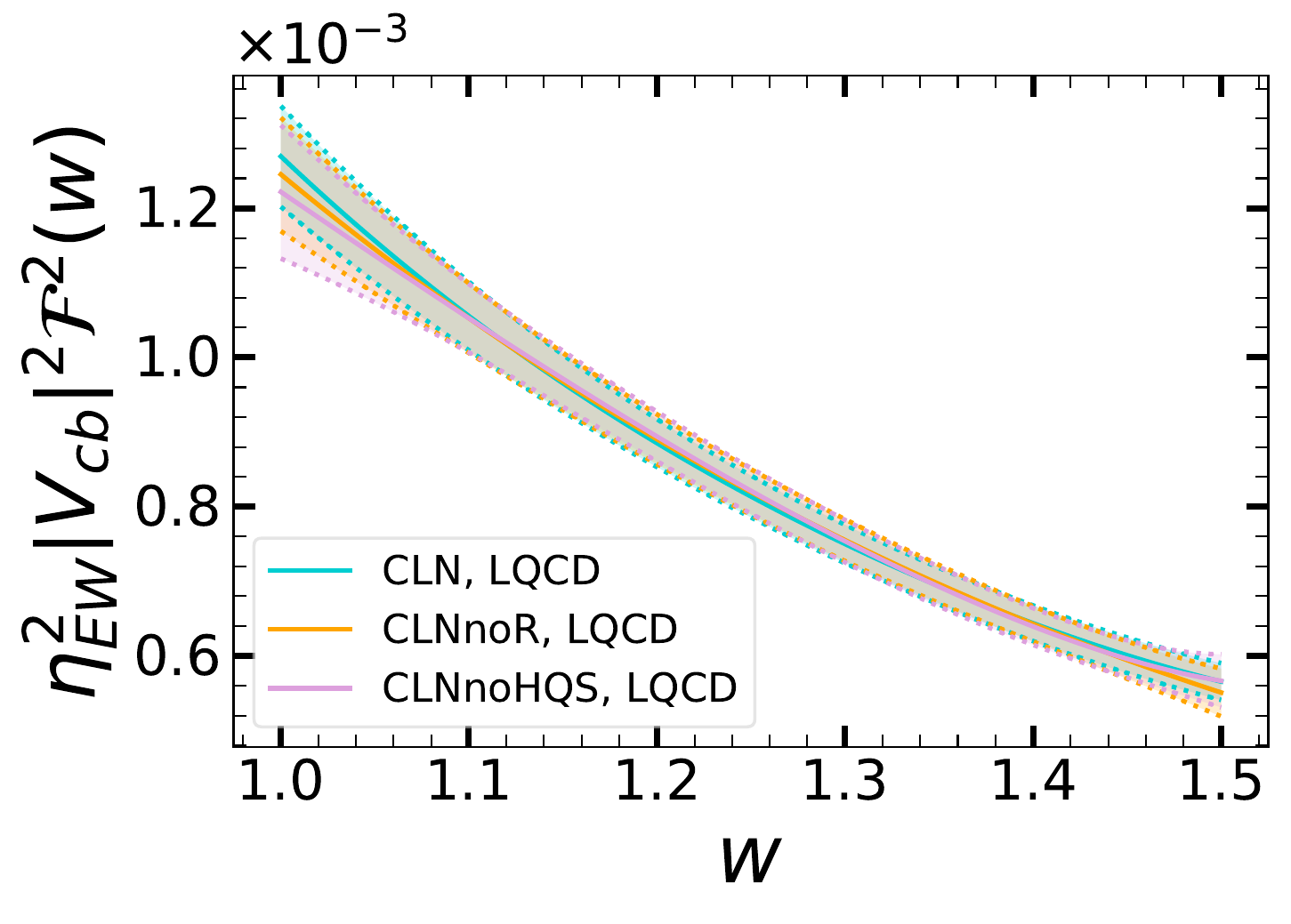}
\includegraphics[width=0.4\linewidth]{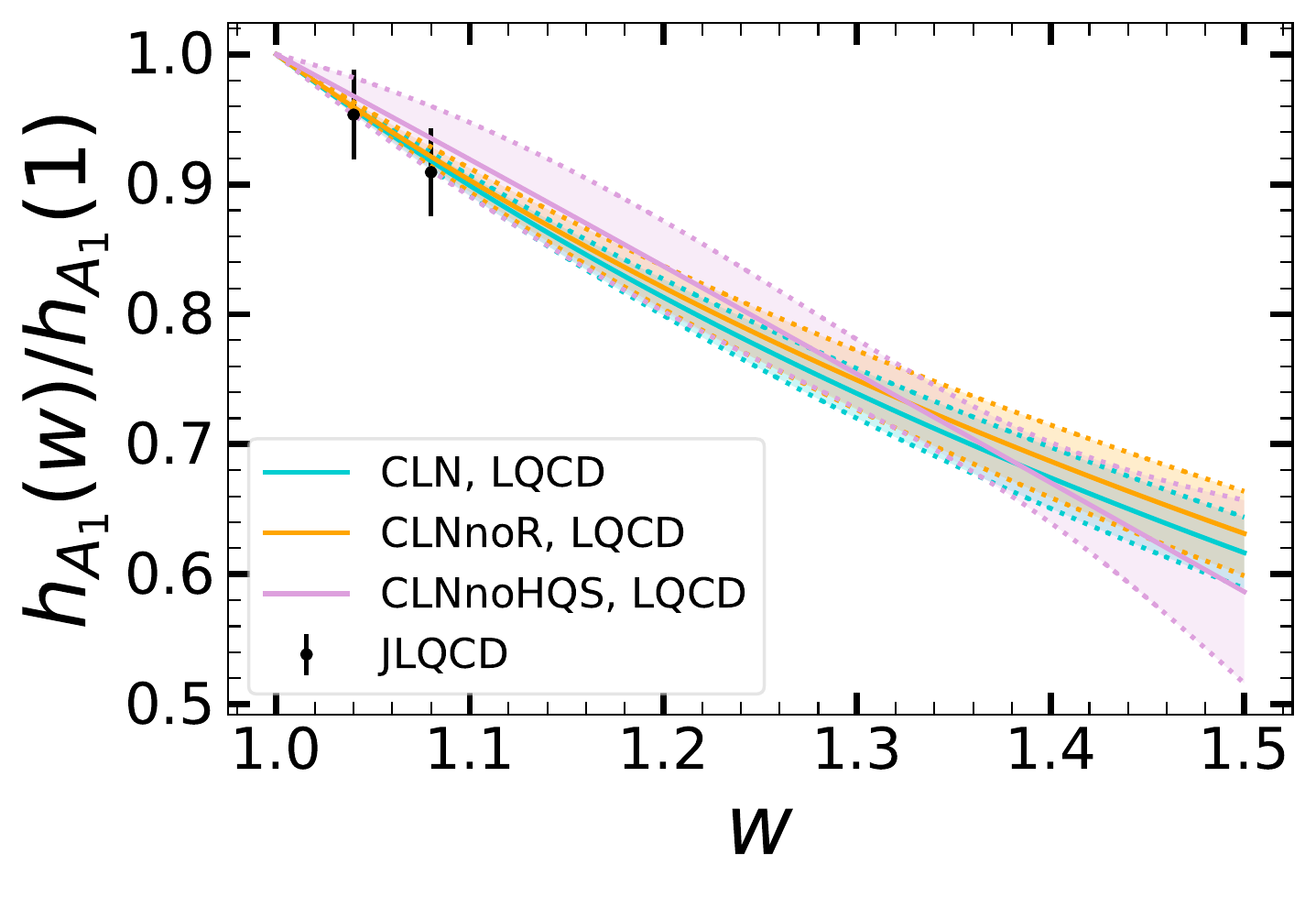}
\includegraphics[width=0.4\linewidth]{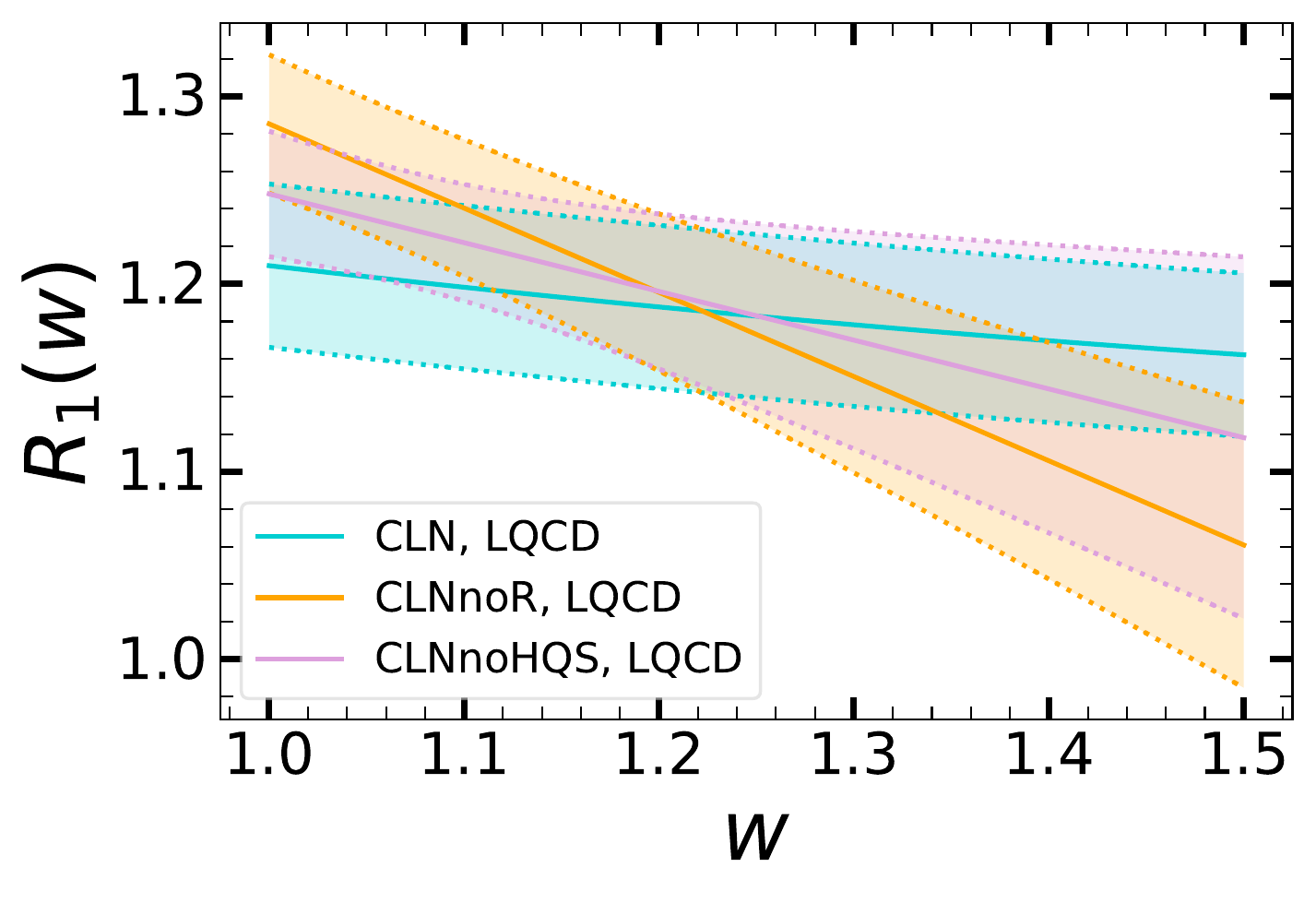}
\includegraphics[width=0.4\linewidth]{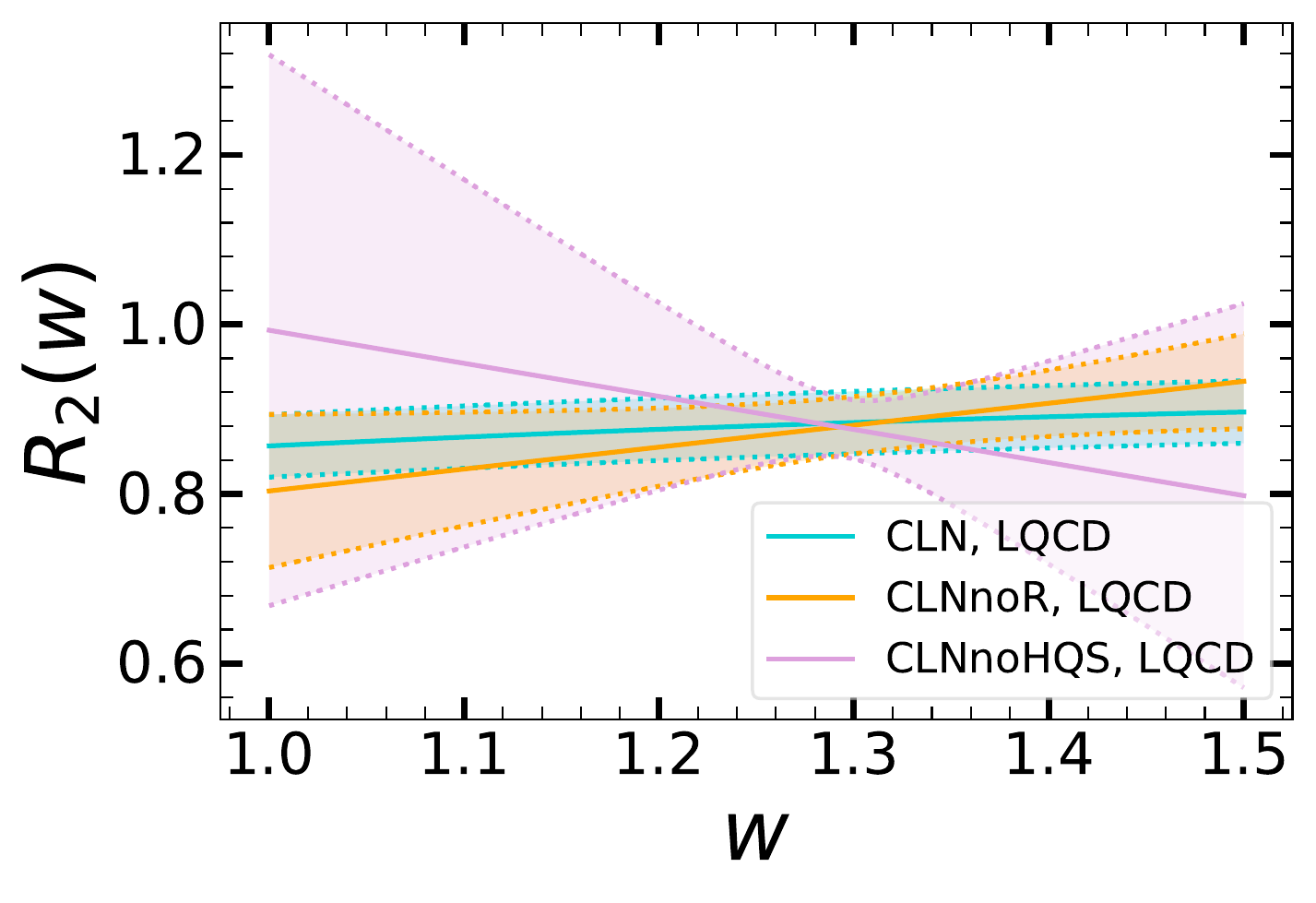}
\caption{The form factors and ratios $\eta_{\rm EW}^2|V_{cb}|^2\mathcal{F}^2$, $h_{A_1}(w)/h_{A_1}(1)$, $R_1$ and $R_2$ as a function of the hadronic recoil, $w$, for CLN, CLNnoR and CLNnoHQS using additional LQCD constraints. The LQCD input has been overlaid in the plot for $h_{A_1}(w)/h_{A_1}(1)$. The central values and uncertainty bands are calculated with the same method as Fig.~\ref{fig:clnfit_noLQCD}.}
 \label{fig:clnfit_LQCD}
 \end{figure*}

A  smaller uncertainty is seen for higher order expansions in BGL with LQCD input, compared to fits without. Figure~\ref{fig:LQCD_hV_eg} emphasizes this, where fits using the BGL parametrization with constraints from LQCD are compatible with the additional $h_V$ data points at non-zero recoil. Therefore, we find that lattice input has brought stability to the higher order BGL configurations.
As discussed in Ref.~\cite{Bernlochner:2019ldg} the BGL(1,0,2) configuration can be considered over-constrained with limited flexibility and therefore not the optimal fit configuration in BGL. Using LQCD constraints, we find little difference between BGL configurations for the CLN-equivalent form factors and form factor ratios at low hadronic recoil, while at larger values, the order of the power series for the three BGL form factors causes a divergence in $R_1$. This is depicted in Figs.~\ref{fig:bglfit_LQCD} and \ref{fig:clnfit_LQCD}, which show a significant reduction in uncertainties when compared to Figs.~\ref{fig:clnfit_noLQCD}-\ref{fig:bglfit_noLQCD}. Similar behavior is seen in the CLNnoR and CLNnoHQS scenarios in $R_1$.

\begin{figure*}
    \centering
    \begin{minipage}{0.4\linewidth}
    \includegraphics[width=\linewidth]{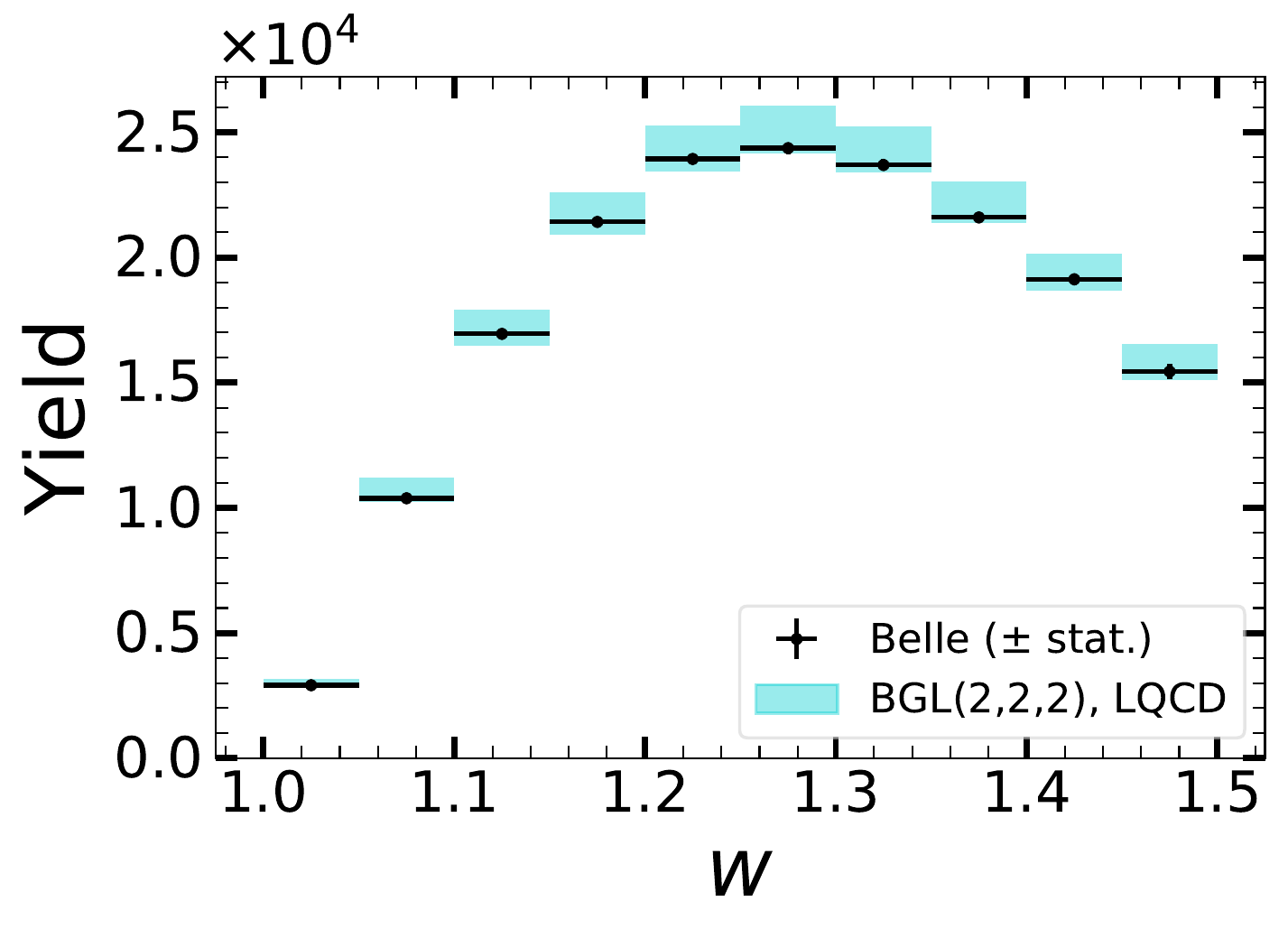}
    \includegraphics[width=\linewidth]{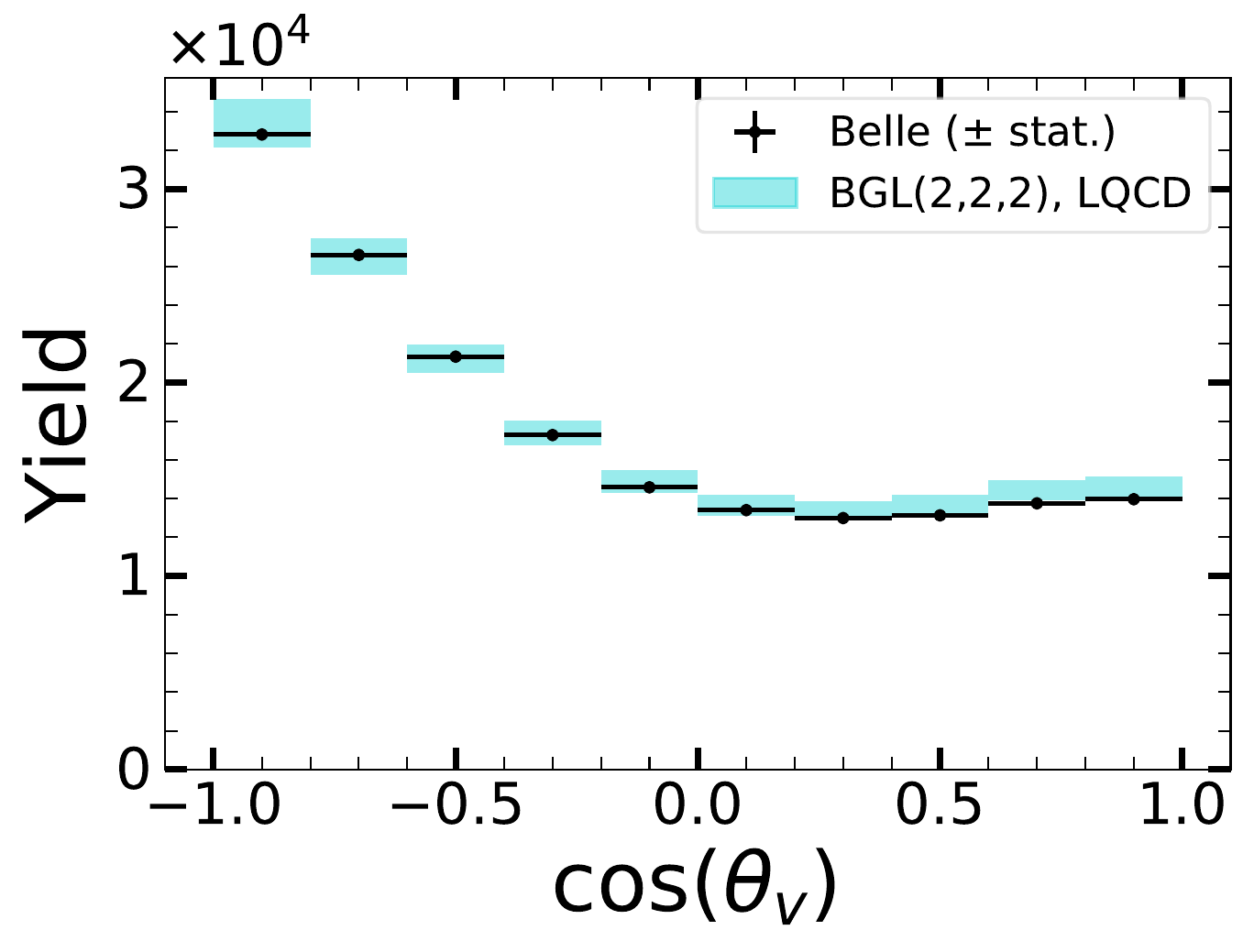}
    \end{minipage}
    \begin{minipage}{0.4\linewidth}
    \includegraphics[width=\linewidth]{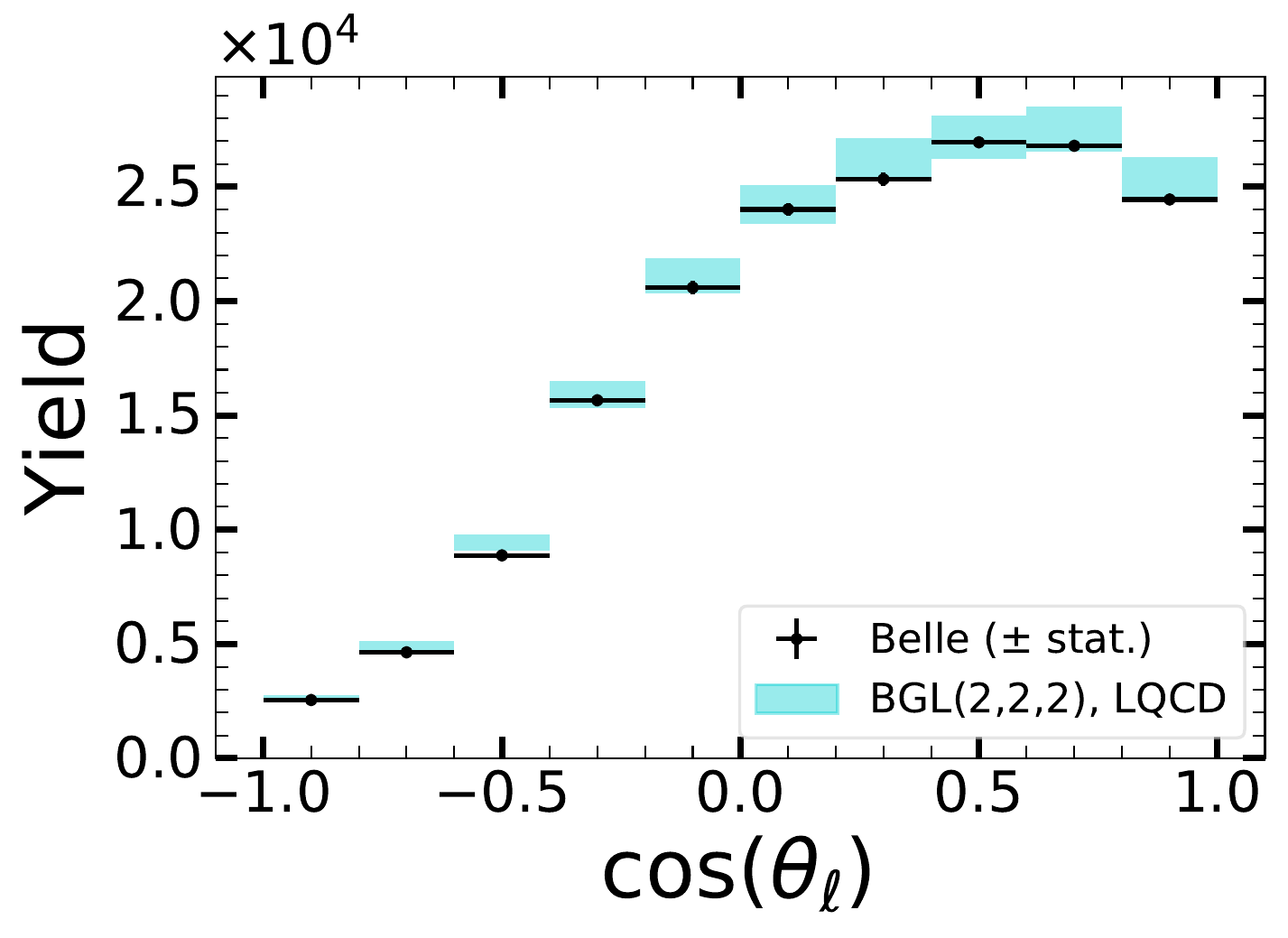}
    \includegraphics[width=\linewidth]{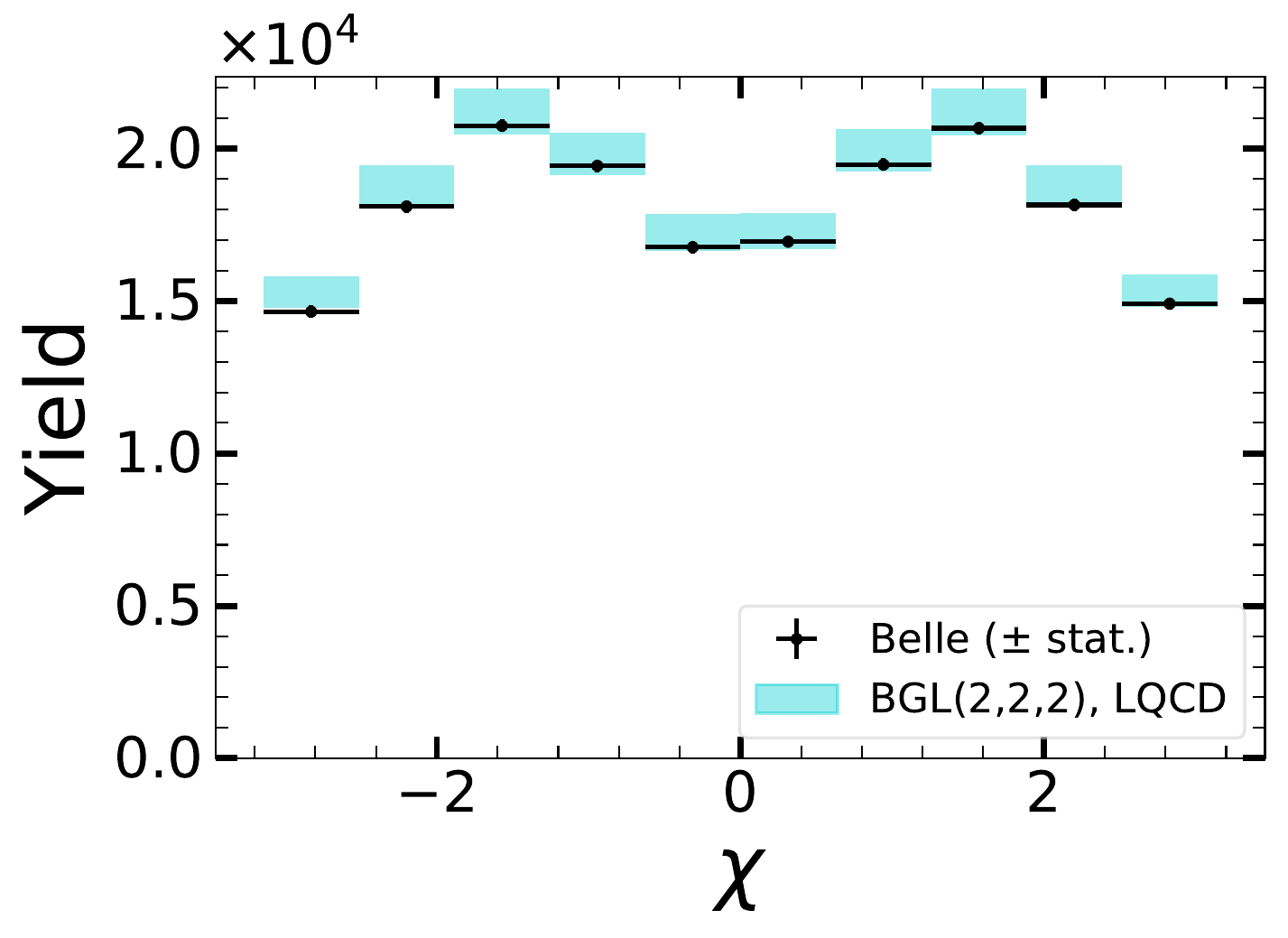}
    \end{minipage}
    \caption{The measured binned yields (data points) with statistical uncertainties for each observable of the $B^{0} \to D^{*-}\ell^+ \nu_\ell$ decay overlaid with the BGL(2,2,2) (cyan) parametrization fit results, using additional LQCD constraints. The statistical and systematic uncertainties in the fits are determined in the same way as Fig.~\ref{fig:both_fits}. The results from the fit are in agreement with the data.}
    \label{fig:fits_LQCD}
\end{figure*}

The inclusion of LQCD constraints has not been found to introduce a bias in the normalization of the fits as the measured $\mathcal{B}(B^{0} \to D^{*-}\ell^+ \nu_\ell)$ values are consistent with previous measurements without LQCD, and the yields, as seen in Fig.~\ref{fig:fits_LQCD}, are compatible with the data.

The results clearly show that obtaining a model-independent value of $|V_{cb}|$ is possible with the use of only four additional constraints from LQCD. 

\subsection{Sensitivity to variation in LQCD inputs}
The LQCD inputs used in this paper are preliminary and subject to change. We therefore explore the sensitivity of $\mathcal{F}(1)\eta_{\rm EW}|V_{cb}|$ to variations in these inputs. Two values were taken at $w = 1.04,1.08$ for each of $h_{A_1}(w)$ and $h_V(w)$, as this was found to be the minimum amount of extra information required for the BGL(2,2,2) parametrization to converge. To measure what effect these values have on the fit results, we establish a toy MC analysis using the standard CLN parametrization. The four non-zero recoil values for this toy MC study are set to
\begin{align}
     h_{A_1}(1.04)/h_{A_1}(1) &= (0.95 \pm 0.05), \nonumber\\
     h_{A_1}(1.08)/h_{A_1}(1) &= (0.91 \pm 0.05), \nonumber\\
     h_{V}(1.04)/h_{V}(1) &= (0.95 \pm 0.05), \nonumber\\
     h_{V}(1.08)/h_{V}(1) &= (0.90 \pm 0.05).
     \label{eq:LQCD_points}
\end{align}
The correlations between these are given in Table~\ref{tab:LQCD_incorr}.
\begin{table*}
 \caption{Correlations between the form factor values at non-zero recoil, $h_{X}(w)/h_{X}(1)$ where $X = A_{1},~V$ and $w = 1.04,1.08$. The entries in this table are approximated from preliminary results from JLQCD. This correlation matrix is multiplied by the uncertainties for each LQCD point and then appended to the $40\times 40$ statistical covariance matrix in a block-diagonal manner.}
\renewcommand{\arraystretch}{1.3}
\begin{tabularx}{0.8\linewidth}{lYYYY } 
\hline \hline
 & $h_{A_1}(1.04)/h_{A_1}(1)$ & $h_{A_1}(1.08)/h_{A_1}(1)$ & $h_{V}(1.04)/h_{V}(1)$ & $h_{V}(1.08)/h_{V}(1)$\\ 
\hline
$h_{A_1}(1.04)/h_{A_1}(1)$ & 1.0 & 0.85 & 0.38 & 0.49\\

$h_{A_1}(1.08)/h_{A_1}(1)$ & 0.85 & 1.0 & 0.18 & 0.44\\

$h_{V}(1.04)/h_{V}(1)$ & 0.38 & 0.18 & 1.0 & 0.93\\

$h_{V}(1.08)/h_{V}(1)$ & 0.49 & 0.44 & 0.93 & 1.0\\
\hline \hline
\end{tabularx}
\label{tab:LQCD_incorr}
\end{table*}
These values have been rounded off from Eq.~\ref{eq:LQCD_points_OG}, with inflated uncertainties to account for the inclusion of any additional systematic errors. The sensitivity to form factor values was explored in two ways; each form factor pair was varied together or each value was varied individually. The inputs were allowed to vary to cover a range of different possible slopes as follows:
\begin{align}
    h_{A_1}(1.04)/h_{A_1}(1) &\in [0.9, 1.0], \nonumber \\
    h_{A_1}(1.08)/h_{A_1}(1) &\in [0.85, 0.95], \nonumber \\
    h_{V}(1.04)/h_{V}(1) &\in [0.9, 1.0], \nonumber \\
    h_{V}(1.08)/h_{V}(1) &\in [0.85, 0.95].
\end{align}
In both treatments of the variation, the range of values obtained for $|V_{cb}|$ changed by at most $0.1\%$ while the largest change over any other fit parameter was $2\%$, which is smaller than the systematic uncertainty calculated in the nominal analysis.

Each correlation was then unfixed individually and set to 100 different values spanning 0 to 1 while the others remained fixed and then the fit was repeated. The fit parameters as a function of each correlation were also found to vary by at most $1\%$, except around finely tuned points where the correlation matrix has eigenvalues nearing zero, at which point the entries in $\mathcal{C}^{-1}$ diverge to infinity. From this we conclude that any results for $|V_{cb}|$ are not highly sensitive to the actual values of the additional form factor constraints at non-zero recoil, under the assumption that the correlation matrix remains invertible.
 
\section{Conclusion} 
\label{sec:Conc}
\begin{figure*}[htb]
    \centering
    \includegraphics[width=0.5\linewidth]{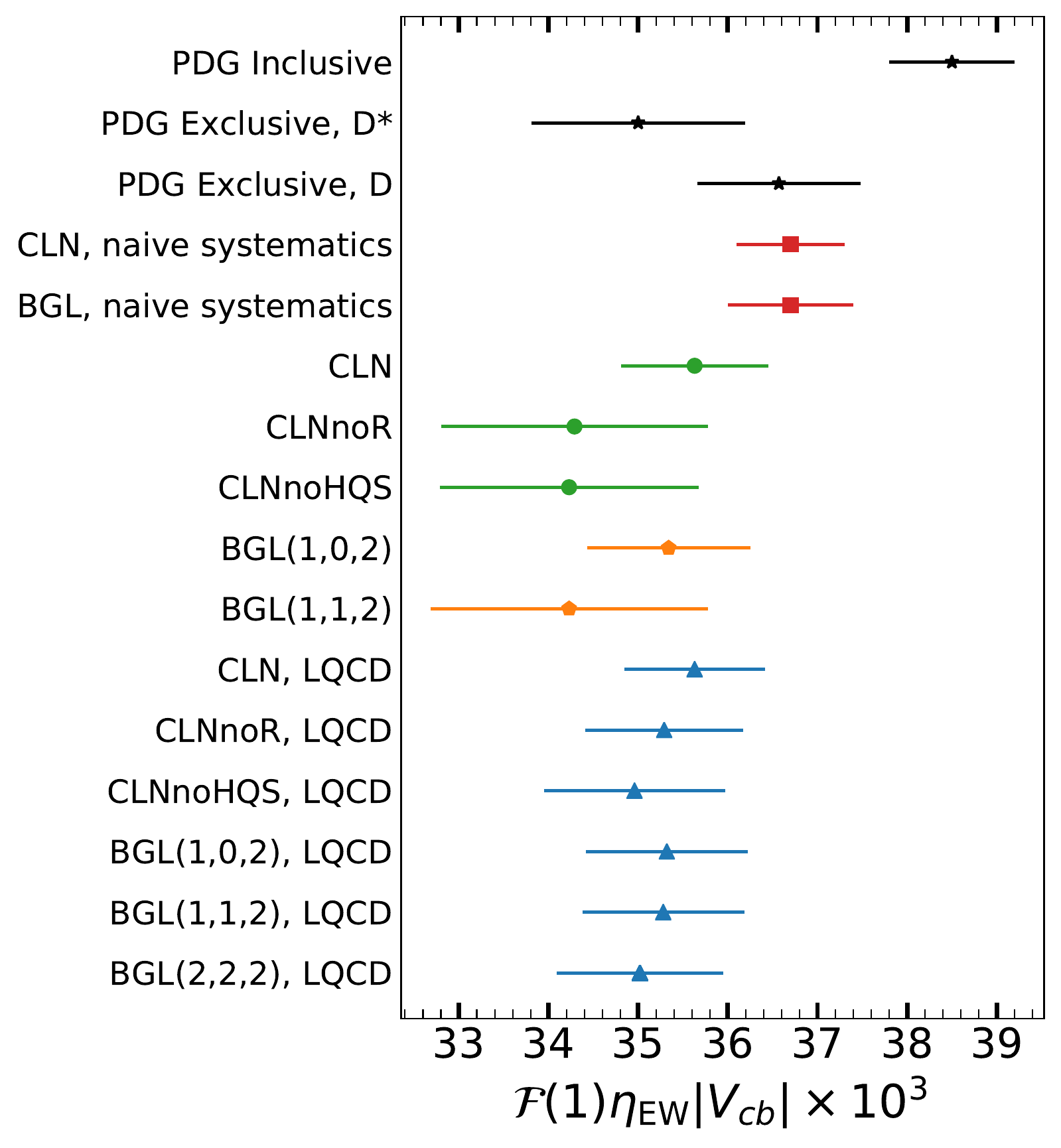}
    \caption{Summary of the $\mathcal{F}(1)\eta_{\rm EW}|V_{cb}|$ values from the various fit scenarios covered in this paper, where the PDG values of $|V_{cb}|$ from Section~\ref{sec:Intro} have been multiplied by the relevant values listed in Table~\ref{tab:constants}.}
    \label{fig:spectrum}
\end{figure*}

We have performed a study of fits to exclusive $B^{0} \to D^{*-} \ell^{+} \nu_{\ell}$ measurements for the determination of $|V_{cb}|$, based on the Belle 2019 untagged measurement. We used preliminary results from the JLQCD group for form factor calculations at non-zero hadronic recoil as additional constraints and find that fits with higher-order parametrizations or less theoretical constraints reliably converge. The $\mathcal{F}(1)\eta_{\rm EW}|V_{cb}|$ results obtained from each of the different methods described in this paper are shown in Fig.~\ref{fig:spectrum}.

We have developed a toy MC approach making use of the covariance properties of the Cholesky decomposition to take into account scale-error dominant systematic uncertainties to avoid bias.
Fits to the CLN parametrization without the requirement of heavy quark symmetry on one or both of the form factor ratios and $h_{A_1}$ have compatible results for the form factors and $\mathcal{F}(1)\eta_{\rm EW}|V_{cb}|$. This demonstrates that the heavy quark symmetry assumptions used in the modelling of CLN are valid and that the results are not indicative of any breakdown in heavy quark symmetry or the use of subleading Isgur-Wise functions.

With the inclusion of JLQCD non-zero recoil constraints we achieve a model-independent result for $\mathcal{F}(1)\eta_{\rm EW}|V_{cb}|$, $(35.02 \pm 0.29 \pm 0.88) \times 10^{-3}$ in BGL(2,2,2) and $(34.96 \pm 0.32 \pm 0.96) \times 10^{-3}$ in CLNnoHQS. Assuming a value of $\mathcal{F}(1)$ from Tab.~\ref{tab:constants} we find $|V_{cb}|=(38.40 \pm 0.32 \pm 0.96) \times 10^{-3}$ and $(38.33 \pm 0.35 \pm 1.05) \times 10^{-3}$, respectively, which remains to be in tension with the inclusive determination.

\section*{Acknowledgements}
We would like to thank T. Kaneko from KEK, Tsukuba, E. Kou from IJCLab, Orsay, and the JLQCD collaboration for their important contributions to lattice calculations and guidance in this analysis. We also thank the Japan Society for the Promotion of Science (JSPS) and the Australian Research Council (ARC DP190101991) for their support. 
\clearpage
\begin{appendices}
\section{Verifying the Cholesky decomposition method}
\label{appendix:A}
\begin{figure*}
 \centering
 \begin{minipage}{0.4\linewidth}
 \includegraphics[width=\linewidth]{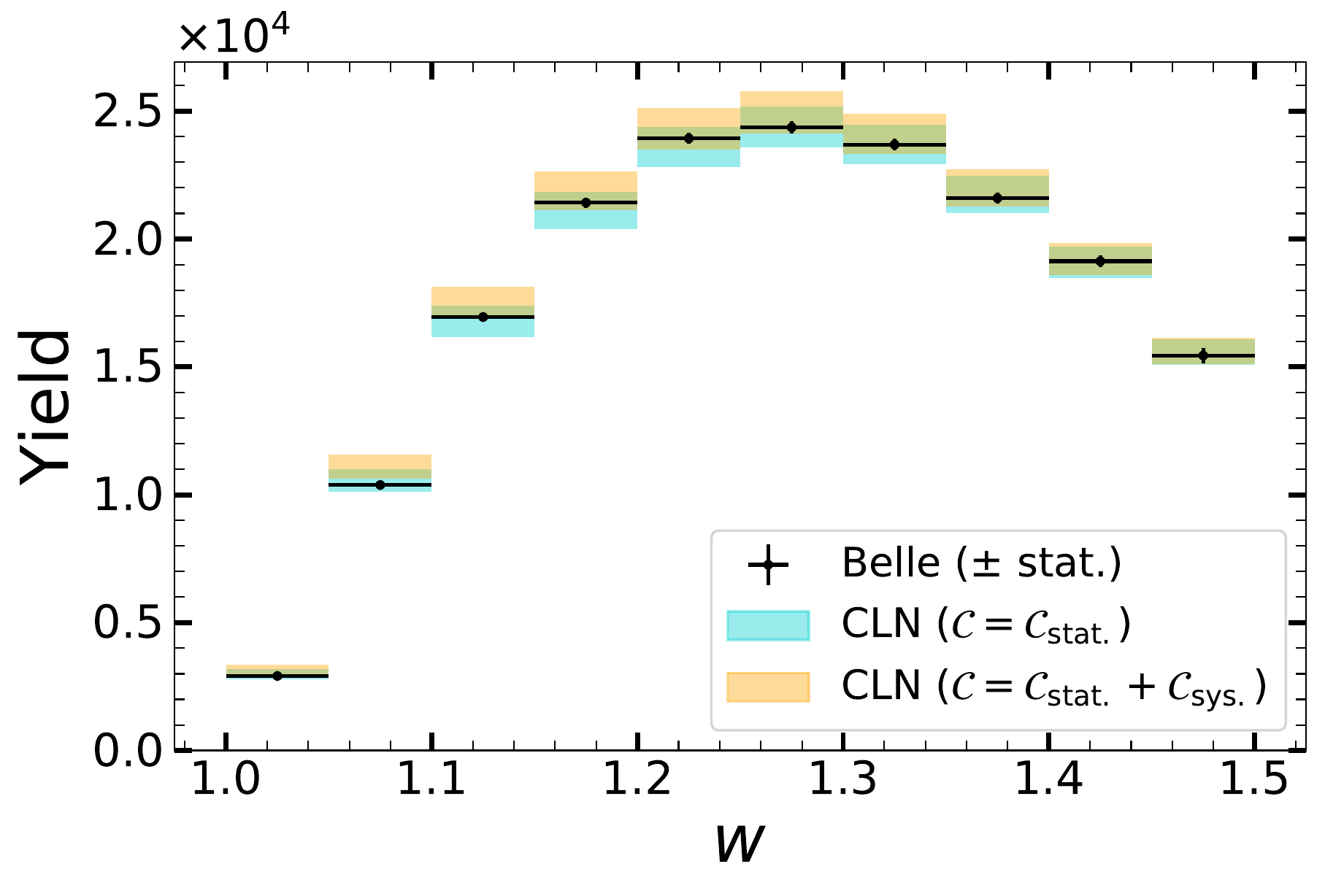}\\
 \includegraphics[width=\linewidth]{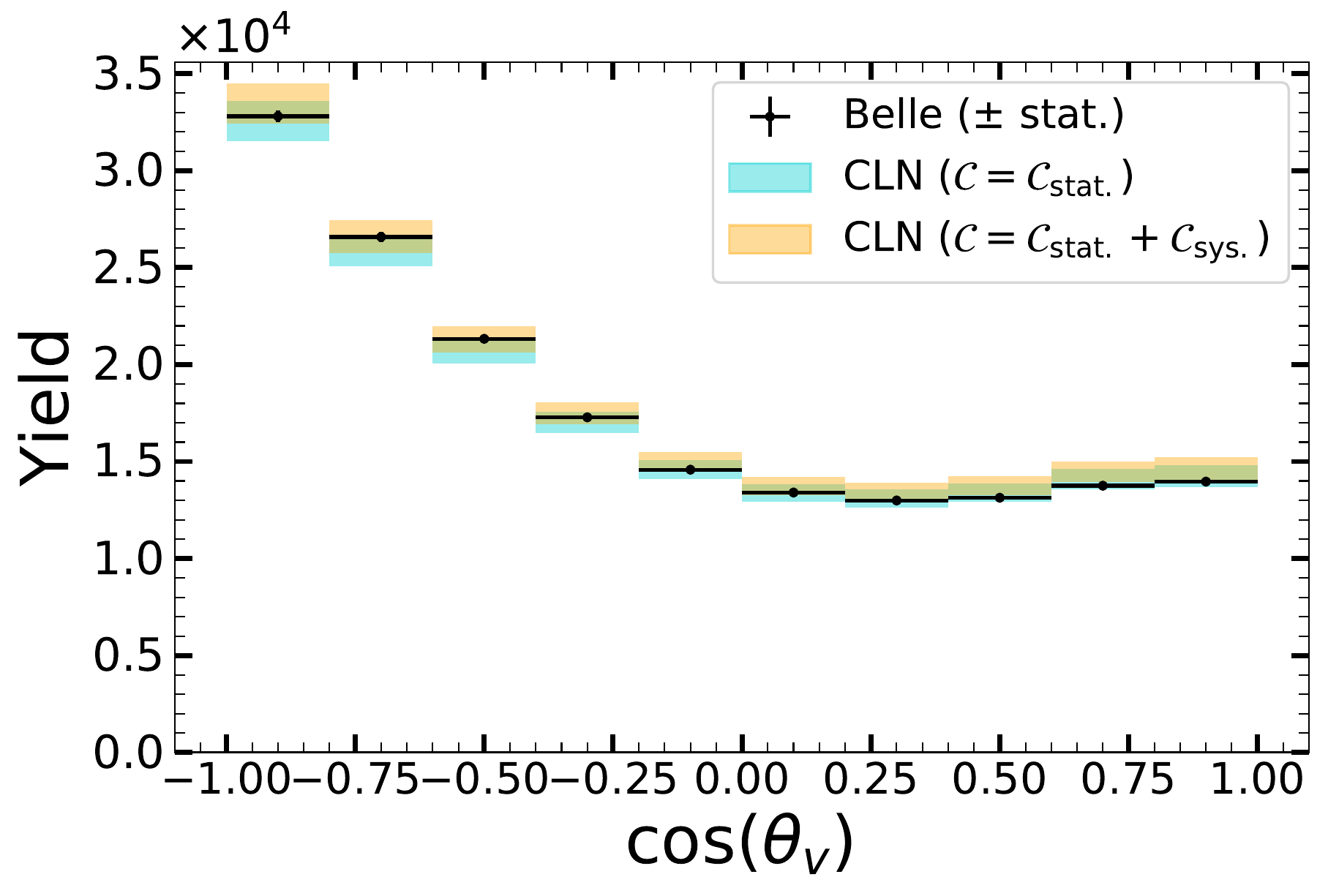}
 \end{minipage}
 \begin{minipage}{0.4\linewidth}
 \includegraphics[width=\linewidth]{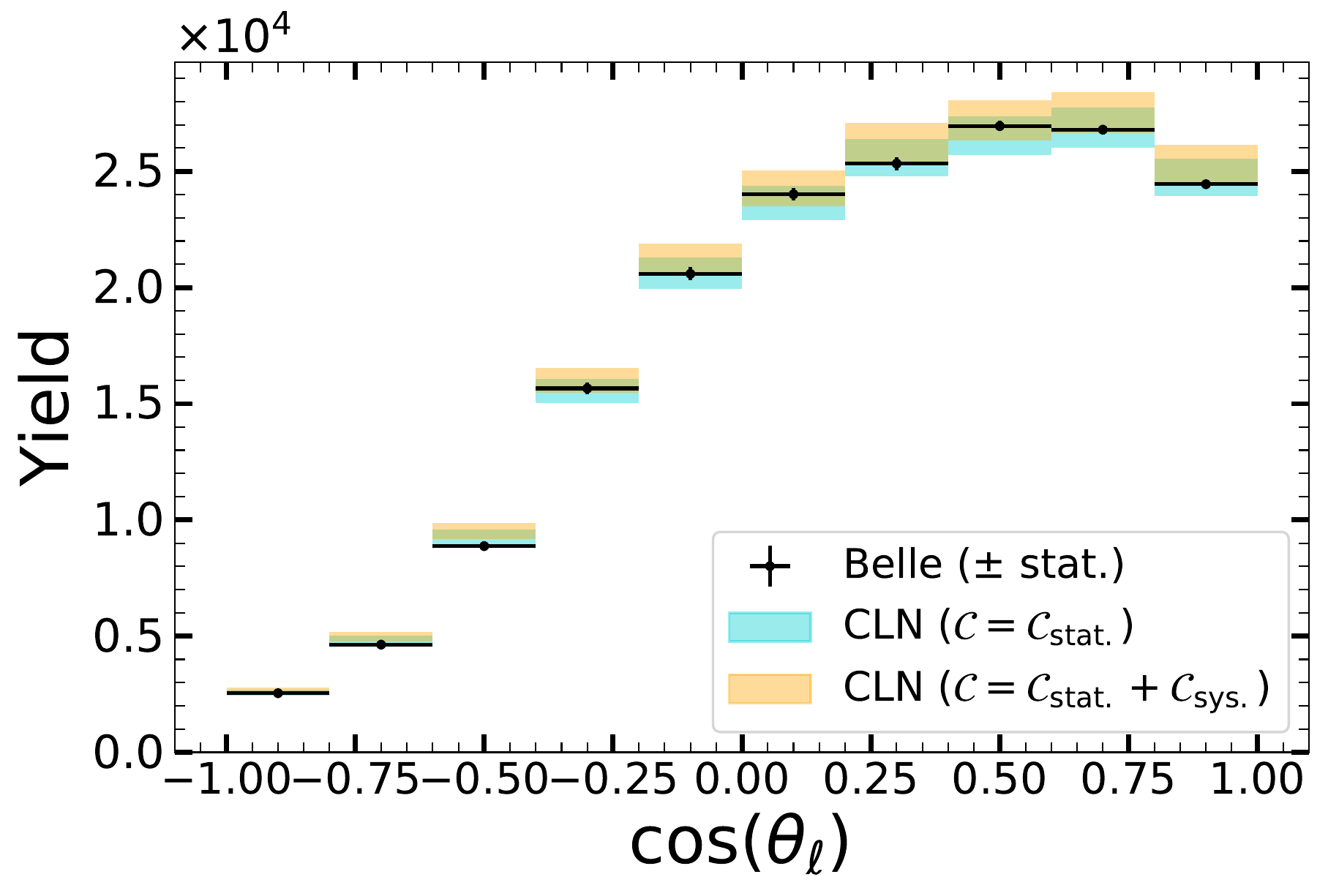}\\
 \includegraphics[width=\linewidth]{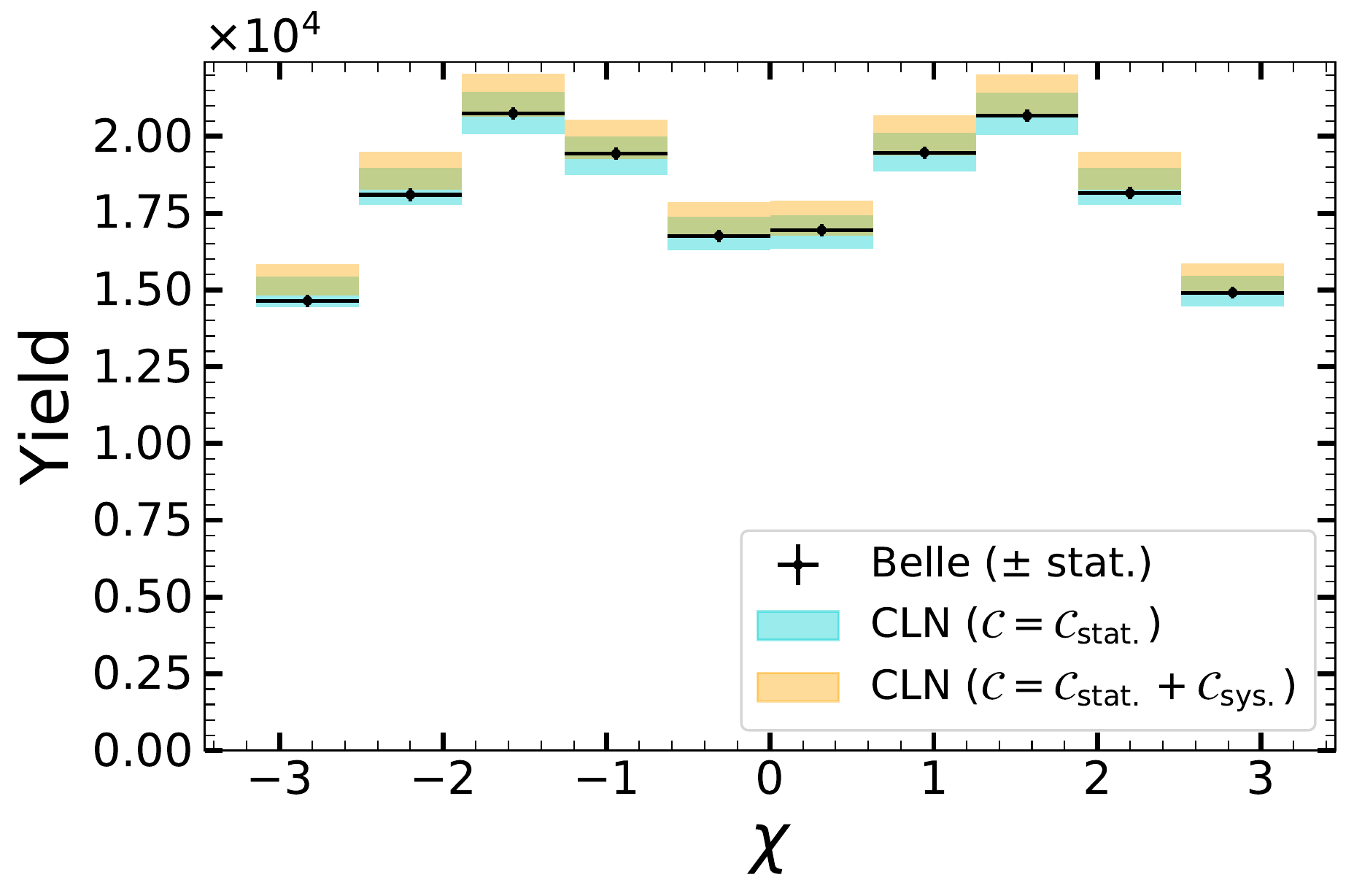}
 \end{minipage}
 
\caption{(Appendix) The measured binned yields (data points) with statistical uncertainty for each observable of the $B^{0} \to D^{*-}\ell^+ \nu_\ell$ decay overlaid with CLN fit results for different covariance matrices, $\mathcal{C}$, used in Eq.~\ref{eq:chi2}. Results are shown for the total covariance matrix being equal to the statistical covariance matrix alone (cyan), and the sum of both the statistical and systematic covariance matrices in a naive way (orange).}
\label{fig:CLN_err}
\end{figure*}

It is well-established that normalization uncertainties for measurements will introduce a bias in $\chi^2$ fits~\cite{DAGOSTINI,cowan1998statistical}. If the systematic uncertainty from the Belle publication is added to the covariance matrix, the minimizing function provides a biased fit due to high correlations and scale-dominant uncertainties, returning a lower $\chi^2$ for expectation values that do not agree with measurements. Figure~\ref{fig:CLN_err} shows a demonstration of this effect. Fit values for the free parameters with and without the addition of the systematic covariance matrix are listed in Table \ref{tab:old_res} and show inconsistency, particularly in the normalization. The data is modelled in a forward-folding approach and this leads to a result that overestimates the yield and therefore also the $\mathcal{B}(B^{0} \to D^{*-}\ell^+ \nu_\ell)$ branching ratio and $\mathcal{F}(1)\eta_{\rm EW}|V_{cb}|$.

\begin{table*}
\caption{(Appendix) Fitted parameters in the CLN and BGL(1,0,2) configurations comparing different total covariance matrices. The results with only statistical uncertainties are obtained by setting $\mathcal{C}$ from Eq.~\ref{eq:chi2} to be equal to $\mathcal{C_{\rm stat.}}$ and the results from combining statistical and systematic uncertainties in a naive way are obtained using $\mathcal{C} = \mathcal{C_{\rm stat.}} + \mathcal{C_{\rm sys.}}$. The branching ratio is obtained from the fit and $\mathcal{F}(1)\eta_{\rm EW}|V_{cb}|$ is calculated from Eq.~\ref{eq:a0f_to_Vcb}.
}
\renewcommand{\arraystretch}{1.3}
\begin{tabularx}{0.8\linewidth}{l  YYYY } 
\hline \hline
Covariance: & stat. & sys.+stat. \\ 
\hline
CLN\\
\hline
$\rho^{2}$ & $1.09 \pm 0.04$ & $1.16 \pm 0.04$\\

$R_{1}(1)$ & $1.20 \pm 0.03$ & $1.18 \pm 0.04$\\

$R_{2}(1)$ & $0.86 \pm 0.02$ & $0.85 \pm 0.03$\\

$\mathcal{F}(1)\eta_{\rm EW}|V_{cb}|\times10^{3}$ & $35.6 \pm 0.2$ & $36.7 \pm 0.6$\\

$\mathcal{B}(B^{0} \to D^{*-}\ell^+ \nu_\ell)$ & $5.04\%$ & $5.18\%$\\

$\chi^{2}$/ndf & $41/36$ & $25/36$\\
\hline
BGL(1,0,2)\\
\hline
$\tilde{a}_0^f \times10^{3}$ & $0.512 \pm 0.004$ & $0.532 \pm 0.010$\\

$\tilde{a}_1^f \times10^{3}$ & $0.6 \pm 0.2$ & $0.2 \pm 0.3$\\

$\tilde{a}_1^{F_1} \times10^{3}$ & $0.30 \pm 0.08$ & $ 0.19 \pm 0.10$\\

$\tilde{a}_2^{F_1} \times10^{3}$ & $-3.8 \pm 1.4$ & $-2.6 \pm 1.8$\\

$\tilde{a}_0^g \times10^{3}$ & $0.93 \pm 0.02$ & $0.93 \pm 0.03$\\

$\mathcal{F}(1)\eta_{\rm EW}|V_{cb}|\times10^{3}$ & $35.3 \pm 0.3$ & $36.7 \pm 0.7$\\
$\mathcal{B}(B^{0} \to D^{*-}\ell^+ \nu_\ell)$ & $5.04\%$ & $5.18\%$\\

$\chi^{2}$/ndf & $39/35$ & $24/35$\\
\hline \hline
\end{tabularx}
\label{tab:old_res}
\end{table*}

To avoid this effect we use a Cholesky decomposition and toy method to propagate systematic uncertainties that are dominated by scale errors. The Belle analysis was repeated with a false data sample generated from arbitrarily chosen CLN parameters:
\begin{eqnarray*}
\rho^2 &=& 1.1,\\
R_{1}(1) &=& 1.2,\\
R_{2}(1) &=& 0.8,\\
\eta_{\rm EW}|V_{cb}| &=& 0.04.
\end{eqnarray*}
The statistical uncertainties in each bin were defined as the square root of the number of events in that bin while the statistical correlations were defined by the following:
\begin{eqnarray}
\text{$\rho_{stat.}$}(i,j) =
\begin{cases} 
      1 & i=j \\
      0.01 & i\neq j \text{, $i$, $j$ block diagonal} \\
      0 & \text{otherwise,} 
   \end{cases}
\end{eqnarray}
where ``$i$, $j$ block diagonal" refers to entries within the $10\times 10$ sub-matrix for each observable along the diagonal of the full $40\times 40$ statistical correlation matrix.

The systematic uncertainty is given by $\sigma_i = N^{exp}_{i}\epsilon_{i}$, where $\epsilon_{i} = 0.01$ in each bin and
\begin{eqnarray}
\text{$\rho_{sys.}$}(i,j) =
\begin{cases} 
      1 & i=j \\
      0.99 & \text{otherwise.} 
   \end{cases}
\end{eqnarray}
The pull obtained from the fits from $10^4$ iterations was measured, where the pull for the $i^{\text{th}}$ fit, $p_i$, of each parameter measurement, $x_i$, is given by
\begin{eqnarray}
p_{i,x} = \frac{x_{i}-\hat{x}}{\sigma_x}.
\end{eqnarray}
Here $\hat{x}$ is the nominal value of the parameter, defined earlier, and $\sigma_x^2 = \sum_i (x_i - \hat{x})^2/n$ for $n$ fits. The results are found to have no bias, with the pull for each fit parameter being consistent with zero. Therefore, we consider this method of producing a toy MC sample via the Cholesky decomposition as suitable for determining the nominal values and uncertainties of parameters in the case of dominant scale errors with high correlations.

\end{appendices}
\bibliography{references}
\end{document}